\documentclass[pdflatex,sn-nature]{sn-jnl}

\usepackage{graphicx}
\usepackage{multirow}
\usepackage{amsmath,amssymb,amsfonts}
\usepackage{amsthm}
\usepackage{mathrsfs}
\usepackage[title]{appendix}
\usepackage{xcolor}
\usepackage{textcomp}
\usepackage{manyfoot}
\usepackage{booktabs}
\usepackage{algorithm}
\usepackage{algorithmicx}
\usepackage{algpseudocode}
\usepackage{listings}
\usepackage{comment}
\usepackage{siunitx}
\theoremstyle{thmstyleone}

\theoremstyle{thmstyletwo}

\theoremstyle{thmstylethree}

\begin{document}

\title[Spectrally optimized implicit neural representations mitigate acquisition-related image quality trade-offs in intracranial 4D Flow MRI]{Spectrally optimized implicit neural representations mitigate acquisition-related image quality trade-offs in intracranial 4D Flow MRI}

\author*[1,7]{\fnm{Oliver} \sur{Welin Odeback}}\email{oliver.welin.odeback@ki.se}
\author[1]{\fnm{Nerea} \sur{González-Aranceta}}
\author[1,7]{\fnm{Sebastián} \sur{Jofre}}
\author[2,7]{\fnm{Javier} \sur{Bisbal}}
\author[3]{\fnm{Edward} \sur{Ferdian}}
\author[4]{\fnm{Alistair A.} \sur{Young}}
\author[1]{\fnm{Pia} \sur{Callmer}}
\author[5]{\fnm{Sameer A.} \sur{Ansari}}
\author[6]{\fnm{Mia} \sur{Bonini}}
\author[6]{\fnm{David} \sur{Nordsletten}}
\author[7]{\fnm{Patrick} \sur{Winter}}
\author[7]{\fnm{Susanne} \sur{Schnell}}
\author[1,8]{\fnm{Tobias} \sur{Granberg}}
\author[1,9]{\fnm{Alexander} \sur{Fyrdahl}}
\author[1,10]{\fnm{David} \sur{Marlevi}}

\affil*[1]{\orgdiv{Dept. Molecular Medicine and Surgery}, \orgname{Karolinska Institutet},
\orgaddress{\street{Karolinska Universitetssjukhuset Solna (L1:00)}, \city{Stockholm}, \postcode{171 76}, \country{Sweden}}}

\affil[2]{\orgdiv{Biomedical Imaging Center}, \orgname{Pontificia Universidad Católica de Chile},
\orgaddress{\street{Av. Vicuña Mackenna 4860, Macul}, \city{Santiago}, \postcode{7820436}, \country{Chile}}}

\affil[3]{\orgdiv{School of Electrical Engineering and Informatics}, \orgname{Bandung Institute of Technology},
\orgaddress{\street{Jl. Ganesha No. 10}, \city{Bandung}, \postcode{40132}, \state{West Java}, \country{Indonesia}}}

\affil[4]{\orgdiv{School of Biomedical Engineering \& Imaging Sciences}, \orgname{King's College London},
\orgaddress{\street{1 Lambeth Palace Rd, South Bank}, \city{London}, \postcode{SE1 7EU}, \country{UK}}}

\affil[5]{\orgdiv{Department of Radiology, Neurology, and Neurological Surgery, Feinberg School of Medicine}, \orgname{Northwestern University},
\orgaddress{\city{Chicago}, \state{IL}, \postcode{60611}, \country{USA}}}

\affil[6]{\orgdiv{Department of Biomedical Engineering}, \orgname{University of Michigan}, \orgaddress{\street{2200 Bonisteel Blvd, 1107 Carl A. Gerstacker Building},
\city{Ann Arbor}, \state{MI}, \postcode{48109-2099}, \country{USA}}}

\affil[7]{\orgdiv{Department of Physics}, \orgname{University of Greifswald},
\orgaddress{\street{Felix-Hausdorff-Str. 6}, \city{Greifswald}, \postcode{174 89}, \country{Germany}}}

\affil[8]{\orgdiv{Department of Neuroradiology}, \orgname{Karolinska University Hospital},
\orgaddress{\street{Hälsovägen 13, O42}, \city{Stockholm}, \postcode{141 86}, \country{Sweden}}}

\affil[9]{\orgdiv{Department of Clinical Physiology}, \orgname{Karolinska University Hospital},
\orgaddress{\street{Eugeniavägen 3, A8:01}, \city{Solna}, \postcode{171 64}, \country{Sweden}}}

\affil[10]{\orgdiv{Institute for Medical Engineering and Science}, \orgname{Massachusetts Institute of Technology},
\orgaddress{\street{45 Carleton St}, \city{Cambridge}, \postcode{02142}, \state{MA}, \country{USA}}}

\abstract{

4D Flow MRI promises unique opportunities for non-invasive mapping of neurovascular hemodynamics, but its clinical utility remains limited by trade-offs between spatiotemporal resolution, acquisition time, and image noise. Deep learning-based super-resolution offers a post-scan solution; however, supervised approaches require paired or domain-specific training data, while patient-specific physics-informed models can be computationally demanding. This creates a need for unsupervised reconstruction models that can adapt to individual acquisitions without paired high-resolution training data, while remaining robust across anatomy, disease severity, and acquisition quality. To address this, we introduce a \emph{spectrally tuned} implicit neural representation (INR) framework for intracranial 4D Flow MRI, representing the velocity field as a continuous coordinate-based neural field using WIRE activation functions with tunable spectral bias. Data-driven and physics-informed variants were evaluated in nine cerebrovascular cases spanning healthy controls and moderate-to-severe intracranial atherosclerotic disease, using CFD-derived high-resolution reference fields, MR-realistic synthetic low-resolution inputs, and direct fitting to acquired in-vivo data. Cohort-optimal spectral-bias settings generalized across subjects with only small increases in relative error compared with subject-specific optima, reducing the need for case-specific hyperparameter tuning. Physics-informed regularization provided modest additional accuracy but required substantially longer optimization, whereas the data-driven INR achieved near-comparable reconstruction performance within minutes. The framework also remained robust to acquisition-realistic noise and moderate VENC-dependent aliasing. These findings establish spectrally optimized INRs as a flexible unsupervised approach for mitigating acquisition-related trade-offs in intracranial 4D Flow MRI.
}


\maketitle

\section{Introduction}\label{sec1}

Quantitative assessment of cerebrovascular hemodynamics is important for understanding neurological disease mechanisms and characterizing altered flow in disorders such as intracranial stenosis, aneurysms, and arteriovenous malformations~\cite{leng2014evaluating, sforza2009hemodynamics, moftakhar2009cerebral}. In this setting, four-dimensional flow magnetic resonance imaging (4D Flow MRI) holds particular promise through its ability to non-invasively image time-resolved, three-directional velocity data across various cardiovascular compartments~\cite{markl20124d}. While presented for general intracranial use~\cite{schnell2017accelerated, jacobs2026cerebrovascular}, extension of 4D Flow MRI into a wider range of intracranial applications remains challenging, with small vessels, tortuous anatomies, and stenosis-related flow accelerations putting high demands on spatial resolution, signal-to-noise ratio, and scan duration, while velocity-encoding (VENC) selection introduces an additional trade-off between velocity sensitivity and the risk of phase-wrapping artifacts~\cite{morgan20214d}. As such, while the technique holds unique abilities in non-invasively mapping intracranial hemodynamics, practical compromises are still made in image quality, acquisition time, or effective resolution, each potentially limiting the broader use of the technique for quantifying regional intracranial flow information~\cite{waahlin20224d, marlevi2021noninvasive, aristova2022accelerated}.

To address these limitations, post-acquisition reconstruction methods, including deep learning-based approaches, have emerged; all aiming at denoising velocity data, improving effective resolution, and recovering flow features that are poorly resolved at standard clinical data settings~\cite{rutkowski2021enhancement, ferdian20204dflownet, shit2022srflow}. Out of numerous approaches, so-called physics-informed neural networks (PINNs) provide a unique way of combining data-driven reconstruction with prior knowledge of the underlying physical system at hand~\cite{karniadakis2021physics}, effectively constraining traditional network approaches by fundamental physiological regularization. For 4D Flow MRI, this holds particular promise where measured velocity data are typically noisy or sparsely resolved, but where underlying governing relationships are well-known (the Navier-Stokes equations). Numerous examples also exist showcasing the utility of PINN-based approaches for phase-contrast MRI, recovering both flow and pressure in various settings~\cite{kissas2020machine, fathi2020super, shone2023deep, kang2025flow}.

However, despite their promise, current PINN formulations still face important challenges, not least for intracranial extensions. So far, most presented models have mainly been applied to simplified, large-vessel, or model-dependent settings, limiting direct applicability to acquired \textit{in-vivo} data. Other presented approaches rely on reduced-dimensional formulations (such as one-dimensional Navier-Stokes models) introducing assumptions that may not hold in the setting of tortuous, multi-bifurcated vascular anatomies. Another
current drawback is that most PINNs are optimized to represent a \emph{single} vascular geometry, requiring explicit retraining for each new subject at hand, severely limiting applicability for wider clinical research use.

As an alternative, so-called implicit neural representations (INRs) - a class of neural networks closely related to PINNs - similarly learn continuous mappings from spatial coordinates to the measured signal~\cite{sitzmann2020implicit, tancik2020fourier, ramasinghe2022beyond}. However, instead of relying on physics-based regularization, the INR field emphasizes the role of network \emph{spectral} properties, which can be tuned to control the balance between fine-scale detail recovery and smoothness
while still allowing computationally efficient patient-specific optimization. Despite strong results in image super-resolution, view synthesis, and 3D shape reconstruction~\cite{park2019deepsdf, chen2021learning, mildenhall2021nerf}, INR-based reconstruction remains largely unexplored in 4D Flow MRI. To date, the main exception is Saitta et al.~\cite{saitta2024implicit}, who demonstrated unsupervised super-resolution and denoising in a localized aortic setting. However, their implementation considered only a single spectral setting, leaving the effect of spectral-bias optimization on reconstruction quality unexplored. Moreover, the potential of INRs for complex cerebrovascular 4D Flow MRI, including stenotic flow, multi-bifurcated anatomy, and generalization across cases, remains unknown.

In this work, we aimed to explore the use of INRs for unsupervised super-resolution of intracranial 4D Flow MRI, focusing in particular on \emph{spectral bias optimization} to enable robust network performance across various intracranial anatomies. Introducing a state-of-the-art Wavelet Implicit neural REpresentation (WIRE)-based INR reconstruction framework with controllable spectral bias~\cite{saragadam2023wire}, we investigate how this bias influences reconstruction accuracy, show how it can be tuned to balance fine-scale detail recovery against noise, and highlight how optimized bias settings generalize across patient cases. By evaluating two complementary variants - one fast data-driven model and one computationally demanding physics-informed extension - we also map out how physics-based regularization and associated training burden impact the possible utility of INRs in the setting of efficient super-resolution intracranial flow imaging. The main contributions of our work lie in \emph{(1)} the novel application of spectral-bias-tuned networks for intracranial 4D Flow MRI, \emph{(2)} the identification of optimal bias tuning for anatomical generalization, and \emph{(3)} the demonstration that data-driven INRs achieve near-comparable performance to physics-informed alternatives at a fraction of the required optimization time. In conjunction, these novelties provide a robust framework for continuous, arbitrary-resolution reconstruction of intracranial 4D Flow MRI, using a uniform spectrally optimized INR setup to allow for broader image-based exploration of regional intracranial flow phenomena in the future.

\begin{figure}[b!]
\centering
\makebox[\textwidth][c]{
    \includegraphics[width=1.2\textwidth]{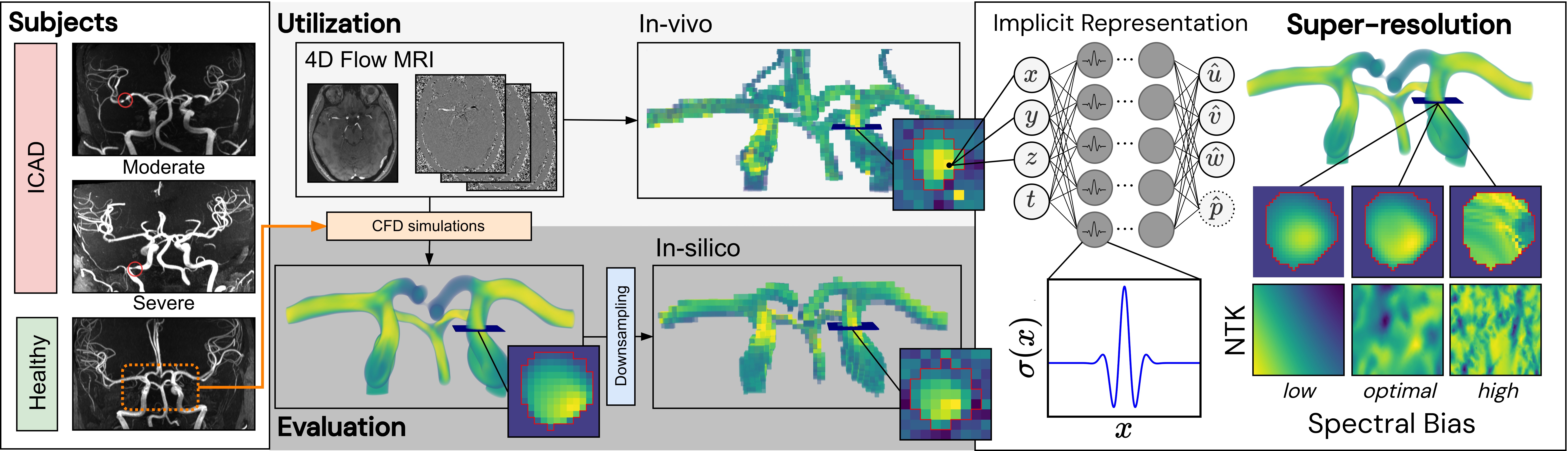}
}
\caption{\textbf{Study overview.} To establish a spectrally optimized implicit neural representation (INR) framework, healthy participants and patients with intracranial atherosclerotic disease (ICAD) underwent cerebrovascular 4D Flow MRI and time-of-flight (TOF) angiography. The INR framework was evaluated by direct \textit{in-vivo} fitting of acquired low-resolution data and quantitative \textit{in-silico} benchmarking using subject-specific CFD simulations. Data-driven and physics-informed INR reconstructions on a super-resolution grid were then evaluated across subjects, with network spectral bias controlled through the tunable activation function $\sigma(x)$ and characterized using neural tangent kernel (NTK) analysis.
}
\label{fig1}
\end{figure}

\section{Methods}\label{sec2}

\subsection{Study design and evaluation overview}\label{sec21}

Fig.~\ref{fig1} summarizes the study design and evaluation workflow. The proposed framework was evaluated in nine cerebrovascular 4D Flow MRI cases, comprising three healthy controls and six patients with intracranial atherosclerotic disease (ICAD), stratified into moderate and severe stenosis groups. Two complementary evaluation settings were used. First, quantitative \textit{in-silico} benchmarking was performed using subject-specific CFD-derived high-resolution reference fields and MR-realistic low-resolution inputs generated through synthetic downsampling, noise injection, and VENC-dependent phase encoding. Second, qualitative feasibility was assessed by fitting the models directly to acquired \textit{in-vivo} 4D Flow MRI data.

Two coordinate-based reconstruction variants were evaluated: a fast data-driven INR optimized against measured velocity data, and a physics-informed INR using the same coordinate-based architecture with additional incompressible Navier--Stokes regularization. Both models used WIRE Gabor activation functions with tunable spectral-bias parameters $(\omega_0,s_0)$. The main analyses assessed reconstruction accuracy, sensitivity to spectral-bias settings, robustness to noise and VENC-dependent aliasing, computational cost, meta-learned initialization, and direct \textit{in-vivo} reconstruction feasibility.

\subsection{Study cohort and MRI acquisitions}\label{sec22}

This retrospective, institutional review board (IRB)-approved study used a sub-cohort from a clinical ICAD protocol at Northwestern Memorial Hospital, where cerebrovascular 4D flow MRI was acquired. We included nine subjects: three healthy volunteers, and six patients with ICAD. The ICAD cases were selected to represent two stenosis-severity groups based on the clinical grading: moderate (50\% stenosis, $n=3$) and severe (70\% stenosis, $n=3$). ICAD stenoses were confirmed from clinical records and MRI/MRA, including vessel wall imaging, by two interventional neuroradiologists.

\begin{table}[t]
\centering
\caption{Subject-level demographics and acquisition characteristics for the included sub-cohort. Voxel size is reported as $(\Delta x,\Delta y,\Delta z)$ in mm. $N_{\mathrm{cardiac}}$ denotes the number of acquired cardiac phases. For ICAD cases, the clinically reported primary lesion location is listed.}
\label{tab:cohort_summary}
\begin{tabular}{llccccccl}
\toprule
\textbf{Group} & \textbf{Case} & \textbf{Sex} & \textbf{Age} & \textbf{Voxel size [mm]} & $N_{\mathrm{cardiac}}$ & \textbf{$\Delta t$ [ms]} & \textbf{Scanner} & \textbf{Stenosis} \\
\midrule
\multirow{3}{*}{Healthy} &
H1 & F & 25 & (0.9821, 0.9821, 1.0) & 7  & 82.6  & Prisma & -- \\
& H2 & M & 31 & (0.9821, 0.9821, 1.0) & 6  & 82.6 & Prisma & -- \\
& H3 & M & 29 & (0.9821, 0.9821, 1.0) & 7  & 82.6 & Prisma & -- \\
\midrule
\multirow{3}{*}{Moderate} &
M2 & M & 35 & (0.9821, 0.9821, 1.0) & 9  & 86.8 & Skyra & RMCA \\
& M1 & M & 61 & (0.9821, 0.9821, 1.0) & 7  & 86.8 & Skyra & RMCA \\
& M3 & M & 65 & (1.0417, 1.0417, 1.0) & 13 & 42.7 & Skyra & BA \\
\midrule
\multirow{3}{*}{Severe} &
S1 & F & 72 & (1.1458, 1.1458, 1.0) & 18  & 42.7 & Skyra & RMCA \\
& S2  & F & 77 & (0.9821, 0.9821, 1.0) & 8  & 86.8 & Skyra & RICA \\
& S3 & F & 74 & (0.9821, 0.9821, 1.0) & 18  & 43.4 & Skyra & LMCA \\
\bottomrule
\end{tabular}
\end{table}

All subjects were scanned at 3T (Siemens Healthineers, Erlangen, Germany) using a Prisma Fit (healthy) or Skyra (patients) system. For all subjects, kt-GRAPPA accelerated (R=5) dual-VENC 4D Flow MRI~\cite{schnell2017accelerated} (low/high VENC = 60/120 cm/s) was acquired. The temporal resolution $\Delta t$ ranged from 42.7-86.8\,ms across subjects, with TE in the range of 3.25-3.52\,ms. The field-of-view was prescribed to cover the circle of Willis and major proximal intracranial arteries. Additionally, a 3D time-of-flight (TOF) gradient-echo sequence (TR/TE = 21-22/3.42\,ms, flip angle = 17$^\circ$) was acquired for segmenting the arteries.

Eddy current correction and noise masking were performed using a semi-automatic MATLAB tool~\cite{schnell2017accelerated}. The same tool was used to co-register the TOF images to the 4D Flow MRI data using rigid-body registration implemented with SPM12 (Statistical Parametric Mapping, version 12)~\cite{spm12}. The resulting aligned TOF images were then used to generate binary segmentation masks for the 4D Flow data.

\subsection{CFD reference simulations}\label{sec23}

To obtain idealized, noise-free high-resolution reference data, we performed patient-specific CFD simulations based on the vascular anatomies and boundary-condition data derived from the cohort described above. Specifically, patient-specific vascular geometries were derived from the co-registered TOF and 4D Flow MRI magnitude images and converted to surface models. After minor surface cleaning and smoothing, inlet and outlet boundaries were defined and the lumen was discretized with an unstructured mesh. Boundary-layer refinement was applied near the vessel wall to resolve laminar near-wall gradients, and an element size of 0.35~mm was used. Inlet boundary conditions were prescribed from 4D Flow-derived velocity using a simplified blunt profile. Outlet flow splits were assigned based on outlet cross-sectional areas in the absence of subject-specific pressure measurements. Blood was modeled as an incompressible Newtonian fluid ($\rho$=1060 kg/m$^3$, $\mu$=0.004 Pa$\cdot$s). Simulations were performed in CHeart~\cite{lee2016multiphysics} with a time step of 0.1 ms and run for four cardiac cycles to reach periodic flow. Nodal output was then mapped onto uniformly voxelized image grids, with isotropic resolution of 0.5 mm and a time discretization of 20 ms to generate the HR velocity data $\mathbf{u}_{\textrm{HR}}$.

\subsection{Synthetic 4D Flow MRI data}\label{sec24}

To generate clinically relevant synthetic LR 4D Flow MRI equivalents $\mathbf{u}_{\mathrm{LR}}$ from the noise-free HR CFD-derived velocity fields $\mathbf{u}_{\mathrm{HR}}$, we used an acquisition-mimicking downsampling pipeline adapted from our previously published dual-VENC framework~\cite{odeback2026potential}. In the present study, the pipeline was modified to use a single-VENC acquisition, corresponding to the high-VENC channel of the previous dual-VENC implementation. This was done to provide a more general benchmark and to allow controlled evaluation of VENC-dependent noise and aliasing effects.

In brief, the HR CFD-derived velocity fields were converted into phase images $\phi$ according to
\begin{equation}
\phi = \left(\frac{\mathbf{u}_{\mathrm{HR}}}{\mathrm{VENC}}\right)\pi.
\label{eq:phase_venc}
\end{equation}
To provide corresponding magnitude information, we used magnitude volumes from \textit{in-vivo} 4D Flow MRI acquired with the same clinical protocol and scanner setup described above (Siemens Magnetom Skyra, prospective k-t GRAPPA dual-VENC acquisition, 120/60~cm/s), obtained from a separate set of five consented subjects from the same IRB-approved study. The magnitude images were aligned to the CFD-derived velocity fields and combined with the synthetic phase images to form complex-valued images,
\begin{equation}
\mathbf{u}_{\mathrm{complex}} = m \cdot e^{i\phi},
\end{equation}
where $m$ denotes the aligned magnitude image.

The complex images were transformed into k-space using a Fast Fourier Transform (FFT). LR equivalents were generated by centrally cropping k-space, corresponding to a two-fold reduction in spatial resolution. For the primary quantitative experiments, complex zero-mean Gaussian noise was then added using a target SNR of 10, chosen to approximate clinically observed noise levels for similar intracranial 4D Flow MRI acquisitions.  The corrupted k-space data were transformed back to image space using an inverse FFT, and velocity images $\mathbf{u}_{\mathrm{LR}}$ were recovered from the phase using the same VENC as in Eq.~\eqref{eq:phase_venc}. The VENC was selected separately for each case to exceed the maximum absolute value across all velocity components in the HR reference, rounded to the nearest 0.1~m/s, with a minimum of 1.2~m/s. This avoided phase wrapping in the main benchmark, and as velocity-to-noise ratio decreases with increasing VENC, case-specific VENC values also introduced differences in effective velocity noise between subjects.

To evaluate temporal super-resolution, the LR velocity fields were additionally downsampled in time by retaining every second time frame and discarding the intermediate frames. Although not fully capturing scanner-level temporal averaging or k-space sampling and reconstruction effects across cardiac phases~\cite{callmer2026deep}, this approach provides a pragmatic and reproducible approximation that preserves direct correspondence between low- and high-resolution time frames.

This yielded a final synthetic input resolution of 1.0~mm spatially and 40~ms temporally.
Table~\ref{tab:synth_params} summarizes the case-specific peak velocities, selected VENC values, temporal dimensions, and fluid-domain voxel counts per timestep used for synthetic 4D Flow MRI generation.

\begin{table}[t]
\centering
\caption{\textbf{Case-specific parameters for synthetic 4D Flow MRI generation.}
All synthetic inputs were generated with target SNR = 10 and two-fold spatial and temporal downsampling.}
\label{tab:synth_params}
\begin{tabular}{lcccccc}
\toprule
\textbf{Case} & $U_{\max}$ [m/s] & VENC [m/s] & $N_{\mathrm{t}}^{\mathrm{HR}}$& $N_{\mathrm{t}}^{\mathrm{LR}}$ & $N_{\mathrm{d}}^{\mathrm{HR}}$ & $N_{\mathrm{d}}^{\mathrm{LR}}$ \\
\midrule
H1 & 1.62 & 1.7 & 25 & 13 & 19,109 & 3,107 \\
H2 & 1.21 & 1.3 & 21 & 11 & 19,371 & 3,185 \\
H3 & 0.99 & 1.2 & 25 & 13 & 20,951 & 3,422 \\
\midrule
M1 & 1.29 & 1.3 & 35 & 18 & 22,672 & 3,687 \\
M2 & 1.28 & 1.3 & 27 & 14 & 14,037 & 2,371 \\
M3 & 5.06 & 5.1 & 26 & 13 & 26,040 & 4,192 \\
\midrule
S1 & 2.58 & 2.6 & 37 & 19 & 13,386 & 2,269 \\
S2 & 4.01 & 4.1 & 31 & 16 & 26,878 & 4,337 \\
S3 & 1.64 & 1.7 & 37 & 19 & 12,019 & 2,042 \\
\bottomrule
\end{tabular}
\end{table}

\subsection{Implicit neural representation model}\label{sec25}

\begin{figure}[b]
\centering
\makebox[\textwidth][c]{
    \includegraphics[width=1.4\textwidth]{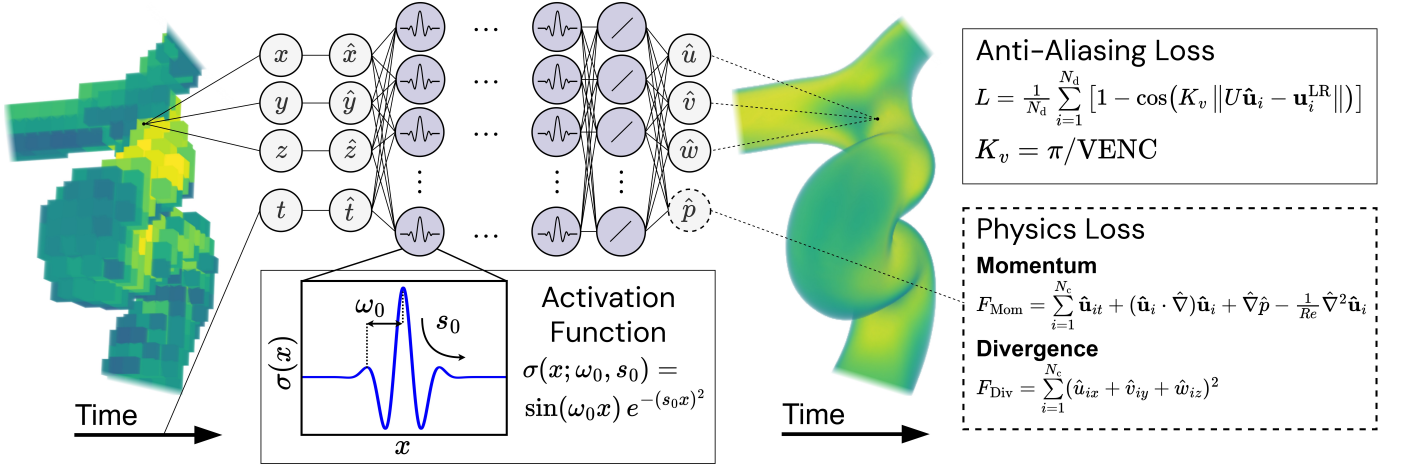}
}
\caption{
\textbf{Coordinate-based INR reconstruction framework.}
The 4D Flow MRI velocity field is represented as a continuous function of spatiotemporal coordinates using a WIRE-based multilayer perceptron with tunable Gabor activation parameters $(\omega_0,s_0)$. The data-driven INR is optimized using only a periodic anti-aliasing loss, while the physics-informed INR uses the same coordinate-based architecture and loss with additional incompressible Navier-Stokes regularization.}\label{fig2}
\end{figure}

We represent the 4D Flow MRI velocity field as a continuous function parameterized by a multi-layer perceptron (MLP). Specifically, an INR is defined as
\begin{equation}
f_{\theta}:\mathbb{R}^{D_{\mathrm{in}}}\rightarrow\mathbb{R}^{D_{\mathrm{out}}},
\end{equation}
where $\theta$ denotes the trainable network parameters. Given an input coordinate $\mathbf{x}\in\mathbb{R}^{D_{\mathrm{in}}}$, where $D_{\mathrm{in}}=4$ for three spatial coordinates and time $(x,y,z,t)$, the INR outputs a vector
\begin{equation}
f_{\theta}(\mathbf{x}) \in \mathbb{R}^{D_{\mathrm{out}}},
\end{equation}
where $D_{\mathrm{out}}=3$ for three velocity components $\hat{\mathbf{u}}(\mathbf{x}) = (\hat{u},\hat{v},\hat{w})$. For an $M$-layer MLP, intermediate activations are computed as
\begin{equation}
\mathbf{y}_{m}=\sigma(\mathbf{W}_{m}\mathbf{y}_{m-1}+\mathbf{b}_{m}), \quad m=1,\dots,M,
\end{equation}
with $\mathbf{y}_0=\mathbf{x}=(x,y,z,t)$ and a final linear readout
\begin{equation}
\hat{\mathbf{u}}(\mathbf{x})=\mathbf{y}_{\mathrm{out}}=\mathbf{W}_{M+1}\mathbf{y}_{M}+\mathbf{b}_{M+1}.
\end{equation}
The choice of activation function $\sigma(\cdot)$ controls the spectral bias of the representation and therefore the balance between recovering fine-scale flow details and suppressing noise. In this work, we adopt the Wavelet Implicit REpresentation (WIRE) framework~\cite{saragadam2023wire}, which controls the spectral bias of an INR using a Gabor activation with two parameters $\omega_0$ and $s_0$. Because the physics-informed model requires repeated automatic differentiation of the network output to evaluate the Navier-Stokes residuals, we use a real-valued Gabor activation rather than the complex-valued formulation in~\cite{saragadam2023wire}, defined as:
\begin{equation}
\sigma(x;\omega_0,s_0)=\sin(\omega_0 x)\,\exp\!\big(-(s_0 x)^2\big),
\label{eq:real_wire_gabor}
\end{equation}
where $\omega_0$ sets the oscillation frequency and $s_0$ controls the width of the Gaussian window. As such, the neural network is formulated as
\begin{equation}
\begin{aligned}
f_{\theta}(\mathbf{x})
&= \mathbf{W}_m (\sigma_{m-1}\circ \cdots \circ \sigma_{0})(\mathbf{x}), \\
\sigma_i(\mathbf{x}_i)
&= \sin{\big(\omega_0 \cdot (\mathbf{W}_i\mathbf{x}_i + \mathbf{b}_i)\big)\,
\exp\!\big(-(s_0 \cdot (\mathbf{W}_i\mathbf{x}_i + \mathbf{b}_i))^2\big)}.
\end{aligned}
\end{equation}
Intuitively, the sinusoidal component controlled by $\omega_0$ promotes the representation of fine-scale, oscillatory flow features, whereas the Gaussian component, controlled by $s_0$, dampens the activation from signals farther from the origin. This localization improves optimization stability and mitigates globally periodic artifacts, such as ringing, that can arise with purely sinusoidal INRs (SIRENs)~\cite{sitzmann2020implicit, saragadam2023wire}. Note that setting $s_0=0$ reduces the activation to the sine nonlinearity used in SIREN, while setting $\omega_0=0$ yields a Gaussian nonlinearity~\cite{ramasinghe2022beyond}.

As input to the network, we assigned each voxel in the LR 4D Flow MRI velocity field $\mathbf{u}_{\textrm{LR}}$ its corresponding spatiotemporal coordinate. To improve numerical stability and avoid large-magnitude inputs during training, the coordinate values were standardized using:
\begin{equation}
\hat{x} = \frac{x - \mu_{\textrm{s}}}{\sigma_{\textrm{s}}} , \hat{y} = \frac{y - \mu_{\textrm{s}}}{\sigma_{\textrm{s}}} ,\hat{z} = \frac{z - \mu_{\textrm{s}}}{\sigma_{\textrm{s}}}, \hat{t}= \frac{t - \mu_{\textrm{t}}}{\sigma_{\textrm{t}}}.
\label{eq:normare}
\end{equation}
Here, $\mu_{\textrm{s}}$ and $\sigma_{\textrm{s}}$ denote the mean and standard deviation computed from a fixed, \emph{template} spatial coordinate axis, chosen to represent a typical field-of-view dimension (20 cm with 1 mm spacing). The same spatial normalization factors were applied to all three spatial axes $(x,y,z)$, ensuring identical scaling of spatial coordinates in the network input. Similarly, $\mu_{\textrm{t}}$ and $\sigma_{\textrm{t}}$ were computed from a fixed template temporal axis (10 s duration with 100 ms spacing). Importantly, this template-based normalization avoids case-specific changes in coordinate scaling caused by differences in field-of-view or resolution, thereby supporting the use of shared spectral-bias hyperparameters $(\omega_0,s_0)$ across datasets.

The target velocity components were normalized by a fixed characteristic velocity $U$:
\begin{equation}
\hat{\mathbf{u}}(\mathbf{x}) = \frac{\mathbf{u}(\mathbf{x})}{U}, \qquad U = 2~\mathrm{m/s}.
\label{eq:vel_norm}
\end{equation}

\subsection{Data fidelity and anti-aliasing loss}\label{sec26}

In the presence of aliased voxels in the MRI data $\mathbf{u}_{\mathrm{LR}}$, a conventional $\ell_2$ loss would severely penalize aliased voxels even when the prediction is physically correct. To make the loss function term consistent with the phase-encoding periodicity and to promote unwrapping of aliased voxels, we employ a periodic (cosine) anti-aliasing loss~\cite{carillo2019unwrapping}, defined as:

\begin{equation}
\mathcal{L}
=\frac{1}{N_{\textrm{d}}}\sum_{i=1}^{N_{\textrm{d}}}
\left[1-\cos\!\Big(K_v\big(U\hat{\mathbf{u}}_{i}-\mathbf{u}_{\mathrm{LR},i}\big)\Big)\right],
\qquad
K_v=\frac{\pi}{\mathrm{VENC}}.
\label{eq:anti_aliasing_loss}
\end{equation}
Because the cosine term is \(2\pi\)-periodic, velocity differences satisfying \(U\hat{\mathbf{u}}_{i}-\mathbf{u}_{\mathrm{LR},i}=2k\,\mathrm{VENC}\), with \(k\in\mathbb{Z}\), yield zero penalty, allowing aliased voxels to be matched modulo the correct velocity while still penalizing deviations unrelated to aliasing.

Herein, \(N_{\textrm{d}}\) denotes the number of data points used to evaluate the loss. We compute the loss only over voxels within the segmented lumen, defined by a binary mask, and exclude background voxels. This restricts optimization to the fluid domain and prevents the network from fitting noise-dominated regions outside the vasculature. As such, application of our framework requires a segmentation mask to delineate the vascular lumen and define the fluid domain over which the loss terms are evaluated.

\subsection{Physics-informed regularization}\label{sec27}

While the data-driven INR can denoise and super-resolve the measured velocity field through the spectral bias of its activation functions, it does not explicitly enforce physical consistency. We therefore introduced a physics-informed variant that regularizes the reconstruction using the incompressible Navier-Stokes momentum and continuity equations. For this model, the network output was extended to include pressure,
\begin{equation}
f_{\theta}(\mathbf{x}) = \big(\hat{u},\hat{v},\hat{w},\hat{p}\big) \in \mathbb{R}^{4},
\end{equation}
where \(\hat{\mathbf{u}}=(\hat{u},\hat{v},\hat{w})\) denotes the normalized velocity field and \(\hat{p}\) the normalized pressure. The pressure output enables the momentum residual to include spatial pressure gradients.

Following~\cite{kissas2020machine, fathi2020super, wang2023expert}, the physics loss was formulated in non-dimensional form using a characteristic length \(L\), characteristic velocity \(U\), and characteristic time \(T=L/U\):
\begin{align}
x^\ast &= \frac{x}{L}, \qquad y^\ast = \frac{y}{L}, \qquad z^\ast = \frac{z}{L}, \qquad t^\ast = \frac{t}{T}, \label{eq:char_coords}\\
\hat{u} &= \frac{u}{U}, \qquad \hat{v} = \frac{v}{U}, \qquad \hat{w} = \frac{w}{U}, \qquad \hat{p} = \frac{p}{\rho U^2}. \label{eq:char_fields}
\end{align}
We fixed \(U=2~\mathrm{m/s}\) and \(L=5~\mathrm{mm}\), corresponding to a representative intracranial artery diameter. Blood was modeled as an incompressible Newtonian fluid with density \(\rho=1060~\mathrm{kg/m^3}\), dynamic viscosity \(\mu=0.004~\mathrm{Pa\cdot s}\), and kinematic viscosity \(\nu=\mu/\rho\).

For the physics-informed INR, coordinate inputs were normalized in two steps. First, spatial and temporal coordinates were non-dimensionalized as defined in Eq.~\eqref{eq:char_coords}. Second, these non-dimensional coordinates were standardized using the template-based normalization described in Eq.~\eqref{eq:normare}. Here, the same spatial and temporal template axes were first divided by $L$ and $T$, respectively. The resulting template means and standard deviations were then used to standardize the non-dimensional input coordinates. This ensured that the physics-informed INR used non-dimensional variables for the Navier-Stokes residuals while preserving an input coordinate distribution comparable to that of the data-driven INR.

With this scaling, the non-dimensional incompressible Navier-Stokes residuals are
\begin{align}
F_{\mathrm{mom}}(\mathbf{x}_i) &=
\frac{\partial \hat{\mathbf{u}}}{\partial t}(\mathbf{x}_i)
+ \big(\hat{\mathbf{u}}(\mathbf{x}_i)\cdot\nabla\big)\hat{\mathbf{u}}(\mathbf{x}_i)
+ \nabla \hat{p}(\mathbf{x}_i)
- \frac{1}{\mathrm{Re}}\nabla^{2}\hat{\mathbf{u}}(\mathbf{x}_i), \label{eq:res_mom}\\
F_{\mathrm{div}}(\mathbf{x}_i) &=
\nabla\cdot\hat{\mathbf{u}}(\mathbf{x}_i),
\label{eq:res_div}
\end{align}
where \(\mathrm{Re}=UL/\nu\). Derivatives were computed by automatic differentiation at collocation points \(\{\mathbf{x}_i\}_{i=1}^{N_{\mathrm{c}}}\) sampled within the fluid domain. The physics-informed loss was defined as
\begin{equation}
\mathcal{L}_{\mathrm{phys}}
=
\frac{1}{N_{\mathrm{c}}}\sum_{i=1}^{N_{\mathrm{c}}}
\left\lVert F_{\mathrm{mom}}(\mathbf{x}_i)\right\rVert_2^2
+
\frac{1}{N_{\mathrm{c}}}\sum_{i=1}^{N_{\mathrm{c}}}
\left(F_{\mathrm{div}}(\mathbf{x}_i)\right)^2.
\label{eq:physics_loss}
\end{equation}

We used \(N_{\mathrm{c}}=1.5\times10^6\) collocation points in all experiments, far exceeding \(N_{\mathrm{d}}\) in all cases. Collocation points were generated from the discrete fluid-domain coordinates used for the data loss. Specifically, coordinates were sampled from the \(N_{\mathrm{d}}\) fluid voxels. To avoid evaluating the PDE residual only at voxel centers, each sampled coordinate was jittered by adding uniform noise drawn from \([-\tfrac{1}{2}\Delta,\tfrac{1}{2}\Delta]\) along each coordinate dimension, where \(\Delta\) denotes the corresponding coordinate spacing. This procedure provides dense, off-grid collocation coverage of the imaged fluid domain.

To balance the data and physics losses during training, the data-loss weight \(\lambda_D\) was updated using the relative gradient magnitudes of the two terms, following the gradient-balancing motivation in~\cite{wang2021understanding} and similar to~\cite{shone2023deep}:
\begin{equation}
\widehat{\lambda}_D^{\,k+1}
=
\frac{\left\lVert \nabla_{\theta}\mathcal{L}_{\mathrm{phys}}\right\rVert_2}
{\left\lVert \nabla_{\theta}\mathcal{L}_{D}\right\rVert_2+\varepsilon},
\qquad
\lambda_D^{k+1}=\alpha\,\lambda_D^{k}+(1-\alpha)\,\widehat{\lambda}_D^{\,k+1},
\label{eq:lambda_update}
\end{equation}
with \(\alpha=0.95\). The total physics-informed loss was therefore
\begin{equation}
\mathcal{L}_{\mathrm{tot}}=\lambda_D\,\mathcal{L}_{D}+\mathcal{L}_{\mathrm{phys}}.
\label{eq:Ltot}
\end{equation}

\subsection{Training and optimization}\label{sec28}

All implicit models were trained using the Adam optimizer with an initial learning rate of $10^{-4}$. For the data-driven INR, training was performed for $8,000$ iterations using mini-batches of $20,000$ randomly sampled data points from the fluid mask. The learning rate was exponentially decayed during training, providing larger updates during early fitting and progressively smaller updates as the reconstruction converged.

All implicit models were trained from random initialization using a fixed random seed. For the real-valued WIRE network, each Gabor layer was initialized using a SIREN-style initialization adapted to the WIRE activation~\cite{sitzmann2020implicit, saragadam2023wire}. First-layer weights and biases were sampled from $\mathcal{U}(-1/d_{\mathrm{in}}, 1/d_{\mathrm{in}})$, whereas hidden-layer weights and biases were sampled from $\mathcal{U}(-\sqrt{6/d_{\mathrm{in}}}/\omega_0, \sqrt{6/d_{\mathrm{in}}}/\omega_0)$, where $\omega_0$ denotes the selected spectral-bias parameter. When $\omega_0=0$, the corresponding scaling was instead defined using $s_0$. The same initialization strategy was used for both the data-driven and physics-informed INR, while meta-learned initialization was used only in the separate experiments described in Section~\ref{sec213}.

For the physics-informed INR, optimization was performed in two stages to account for the increased complexity of the combined data and physics objective. The model was first trained with Adam for $10,000$ iterations using mini-batches of $6,000$ data points and $6,000$ collocation points per iteration. It was then refined using L-BFGS~\cite{liu1989limited} for an additional $5,000$ iterations, with a learning rate of $5\times10^{-2}$ and a strong-Wolfe line search. This Adam-to-L-BFGS strategy is commonly used in PINNs, as Adam provides a robust warm start, while L-BFGS can improve convergence once the solution is sufficiently close to the global optimum~\cite{cuomo2022scientific}.

Our INR framework was implemented in PyTorch (v2.2.1), and all training was performed on an NVIDIA A100 GPU with 40 GB memory.

\subsection{Evaluation metrics}\label{sec29}

After training, each model was evaluated by querying the learned continuous representation on the HR spatiotemporal coordinate grid, and comparing the predicted velocity field $\mathbf{u}_{\mathrm{pred}}$ voxel-by-voxel against the HR reference $\textbf{u}_{\textrm{HR}}$ within the fluid mask. We report the mean absolute error (MAE), the directional error (DE), mean relative error (MRE), and the velocity normalized root-mean-square error (vNRMSE), defined as below:

\begin{equation} \begin{aligned} \text{MAE} &= \frac{1}{N_{\textrm{d}}} \sum_{i=1}^{N_{\textrm{d}}} \left\| \mathbf{u}_{\text{pred}}^i - \mathbf{u}_{\text{HR}}^i \right\|_2 \\
\text{DE}\,[\%] &= 100\cdot\bigg(\frac{1}{N_{\textrm{d}}} \sum_i^{N_{\textrm{d}}}\left(1 - \frac{\left| \mathbf{u}_{\text{pred}}^i \cdot \mathbf{u}_{\text{HR}}^i \right|}{\left\| \mathbf{u}_{\text{pred}}^i \right\|_2 \cdot \left\| \mathbf{u}_{\text{HR}}^i \right\|_2 + \epsilon} \right)\bigg) \\
\text{MRE}\,[\%] &= 100\cdot\bigg(\frac{1}{N_{\textrm{d}}} \sum_{i=1}^{N_{\textrm{d}}} \tanh\left(\dfrac{\left\| \mathbf{u}_{\text{pred}}^i - \mathbf{u}_{\text{HR}}^i \right\|_2}{\left\| \mathbf{u}_{\text{HR}}^i \right\|_2 + \epsilon}\right)\bigg) \\
\text{vNRMSE} &= \frac{1}{\max_i{\left\| \mathbf{u}_{\text{HR}}^i \right\|_2}} \sqrt{\frac{1}{N_{\textrm{d}}}\sum_{i=1}^{N_{\textrm{d}}} \left\| \mathbf{u}_{\text{pred}}^i - \mathbf{u}_{\text{HR}}^i \right\|^2_2}
\end{aligned} \label{eq:error_metrics} \end{equation}
 where $\mathbf{u}_{\text{pred}}^i$ and $\mathbf{u}_{\text{HR}}^i$ denote the super-resolved and HR velocity vectors at voxel $i$, respectively, $\epsilon=10^{-6}$, and $N_{\textrm{d}}$ is the total number of fluid-containing voxels.

\subsection{Comparator methods}\label{sec210}

To benchmark the proposed implicit reconstruction framework against supervised convolutional approaches, we compared the INR models with two pre-trained deep-learning baselines. The first was cerebrovascular 4DFlowNet~\cite{ferdian20204dflownet}, using publicly available weights from \url{https://github.com/EdwardFerdian/4DFlowNet}. This model was trained using a synthetic data-generation strategy that differs from the downsampling pipeline used in the present study. The second was GAN-Gen~\cite{odeback2026potential}, using publicly available weights from \url{https://github.com/oliverwelinodeback/4DFlowGAN}. GAN-Gen was trained using the same MR-realistic downsampling pipeline as in the present work, but in a dual-VENC reconstruction setting rather than the single-VENC setting evaluated here. Both convolutional baselines perform spatial super-resolution by a factor of two but do not perform temporal super-resolution. Therefore, CNN-based predictions were evaluated against the HR reference at the corresponding acquired temporal frames, whereas the INR-based methods were queried on the full HR spatiotemporal grid.

\subsection{Spectral bias selection, robustness analyses, and in-vivo fitting}\label{sec211}

To assess sensitivity to spectral bias and identify settings that generalized across anatomies, we systematically varied the WIRE hyperparameters $(\omega_0,s_0)$ over the range 0-100 with a stride of 10, similar to Saragadam et al.~\cite{saragadam2023wire}. For each configuration, models were trained using the full optimization protocol and evaluated using end-of-training voxel-wise error against the HR reference. For the physics-informed INR, performance was evaluated after Adam-L-BFGS optimization; however, when L-BFGS refinement produced numerical non-convergence or increased the reconstruction error relative to the Adam endpoint, the Adam endpoint was retained. Isolated configurations with clear non-convergence were repeated with a different random seed. This convergence-handling procedure was applied before selecting the cohort-optimal spectral bias settings and was used only to exclude failed optimization runs.

To evaluate robustness to acquisition degradation, we performed two additional analyses in three representative cases: one healthy control, one moderate ICAD case, and one severe ICAD case. First, we varied the injected SNR during synthetic LR generation to quantify sensitivity to measurement noise. Second, we varied the VENC parameter in the downsampling pipeline to induce different degrees of phase wrapping and evaluate anti-aliasing performance. Finally, to demonstrate applicability to acquired data, the INR models were fitted directly to representative \textit{in-vivo} 4D Flow MRI cases and evaluated qualitatively.

\subsection{Empirical neural tangent kernel visualization}\label{sec212}

To visualize inductive bias differences between INR configurations, we adopted the approach of Y\"uce et al.~\cite{yuce2022structured} and visualized the empirical NTK at initialization.
Specifically, to characterize the inductive bias induced by different WIRE spectral bias settings, we analyzed the empirical NTK of the initialized network. For a network
$f_{\theta}: \mathbb{R}^{d}\rightarrow\mathbb{R}^{m}$ with parameters $\theta$, the empirical NTK between two input coordinates $\mathbf{x}_i$ and $\mathbf{x}_j$ is defined as
\begin{equation}
K(\mathbf{x}_i,\mathbf{x}_j)
=
\nabla_{\theta} f_{\theta}(\mathbf{x}_i)^{\top}
\nabla_{\theta} f_{\theta}(\mathbf{x}_j),
\label{eq:ntk_pairwise}
\end{equation}
where $\nabla_{\theta} f_{\theta}(\mathbf{x})$ denotes the Jacobian of the network output with respect to all trainable parameters. For a set of $N$ input coordinates, this yields an $N\times N$ kernel matrix
\begin{equation}
K_{ij}
=
\left\langle
J_{\theta}(\mathbf{x}_i),
J_{\theta}(\mathbf{x}_j)
\right\rangle,
\label{eq:ntk_matrix}
\end{equation}
where $J_{\theta}(\mathbf{x}_i)$ is the flattened parameter Jacobian at coordinate $\mathbf{x}_i$. Since the NTK depends on the network architecture, activation function, and initialization, it provides a diagnostic of the frequency-dependent inductive bias before training.

The empirical NTK was computed at initialization for selected spectral bias settings. To enable direct visualization, input coordinates were sampled on a regular two-dimensional grid corresponding to an axial spatial slice, while the remaining spatial coordinate and time coordinate were fixed. Specifically, for the four-dimensional input coordinate $\mathbf{x}=(t,x,y,z)$, we varied $(x,y)$ on a $64\times64$ grid and fixed $t$ and $z$ to the center of their normalized coordinate ranges, yielding $N=4096$ input points. This produced a two-dimensional cross-section of the four-dimensional NTK.

Parameter Jacobians were computed using automatic differentiation in PyTorch. To avoid explicit Python loops over all input coordinates, we used vectorized Jacobian computation with \texttt{torch.func} operations, evaluating per-sample Jacobians in batches. For our multi-output velocity networks, the Jacobians were reduced over output channels to obtain a scalar NTK matrix. The full NTK matrix was assembled block-wise to reduce memory usage:
\begin{equation}
K =
\begin{bmatrix}
K_{11} & \cdots & K_{1B}\\
\vdots & \ddots & \vdots\\
K_{B1} & \cdots & K_{BB}
\end{bmatrix},
\label{eq:ntk_blocks}
\end{equation}
where each block was computed from mini-batches of input coordinates.

The resulting symmetric kernel matrix was decomposed as
\begin{equation}
K \mathbf{v}_{k} = \lambda_k \mathbf{v}_{k},
\label{eq:ntk_eigendecomposition}
\end{equation}
and the dominant eigenvectors were used to visualize the spatial modes most strongly emphasized by the initialized network. Because the NTK coordinates were sampled on a regular grid, each eigenvector $\mathbf{v}_{k}\in\mathbb{R}^{4096}$ was reshaped directly into a $64\times64$ image without interpolation. These eigenfunction visualizations were used to compare low-, intermediate-, and high-spectral bias configurations at initialization, using both SIREN and WIRE architectures.

\subsection{Meta learning}\label{sec213}

To reduce the number of optimization steps required for patient-specific adaptation, we evaluated a meta-learned initialization for both the data-driven and the physics-informed INR. We used Model-Agnostic Meta-Learning (MAML)~\cite{finn2017model} to learn a shared set of initial parameters $\theta^\ast$ across subjects, such that a small number of gradient updates from $\theta^\ast$ could rapidly adapt the network to a new case.

Each subject was treated as one task $\tau_i$. Six subjects were used for meta-training, with two subjects from each severity group (H1-2, M1-2, S1-2), and three subjects were held out for validation (H3, M3, S3). At each meta-iteration, the low-resolution fluid-domain voxels of each sampled task were randomly divided into a support set $\mathcal{S}_i$ and a query set $\mathcal{Q}_i$. The support set was used for task-specific adaptation in the inner loop, whereas the query set was used to update the shared initialization in the outer loop.

The MAML objective was formulated as
\begin{equation}
\theta^\ast =
\arg\min_{\theta}
\mathbb{E}_{\tau_i}
\left[
\mathcal{L}_{\mathrm{outer}}
\left(
\tilde{\theta}_i;\mathcal{Q}_i,\mathcal{C}_i
\right)
\right],
\label{eq:maml_objective}
\end{equation}
where $\tilde{\theta}_i$ denotes the task-adapted parameters obtained after $K$ inner-loop gradient steps from the current initialization $\theta$, and $\mathcal{C}_i$ denotes collocation points used when physics-informed meta-learning was applied. The inner-loop update for task $\tau_i$ was
\begin{equation}
\theta_{k+1}^{(i)}
=
\theta_k^{(i)}
-
\alpha_{\mathrm{inner}}
\nabla_{\theta_k^{(i)}}
\mathcal{L}_{\mathrm{data}}
\left(
\theta_k^{(i)};\mathcal{S}_i
\right),
\qquad
k=0,\ldots,K-1,
\label{eq:maml_inner_update}
\end{equation}
with $\theta_0^{(i)}=\theta$. We used $K=10$ inner-loop steps with learning rate $\alpha_{\mathrm{inner}}=6.11\times10^{-4}$. The meta-parameters were updated using Adam with learning rate $\alpha_{\mathrm{outer}}=1.51\times10^{-5}$ for $10,000$ meta-iterations. Gradient norms were clipped to $\lVert\nabla_\theta\mathcal{L}_{\mathrm{meta}}\rVert_2 \leq 1.0$ for numerical stability.

For the data-driven INR, the anti-aliasing data loss was used in both the inner and outer loops:
\begin{equation}
\mathcal{L}_{\mathrm{outer}}
\left(
\tilde{\theta}_i;\mathcal{Q}_i
\right)
=
\mathcal{L}_{\mathrm{data}}
\left(
\tilde{\theta}_i;\mathcal{Q}_i
\right).
\label{eq:maml_outer_data}
\end{equation}
Thus, the resulting initialization encoded a data-driven prior over cerebrovascular velocity fields while remaining adaptable to each new subject through gradient-based fine-tuning.

For the physics-informed INR, including the Navier-Stokes loss directly in the inner loop was computationally infeasible as the physics residual requires higher-order automatic differentiation, and MAML differentiates through the inner-loop updates. To avoid this nested higher-order differentiation, the inner loop was kept data-driven, while the physics loss was applied in the outer loop after task adaptation:
\begin{equation}
\mathcal{L}_{\mathrm{outer}}
\left(
\tilde{\theta}_i;\mathcal{Q}_i,\mathcal{C}_i
\right)
=
\mathcal{L}_{\mathrm{data}}
\left(
\tilde{\theta}_i;\mathcal{Q}_i
\right)
+
\lambda
\mathcal{L}_{\mathrm{phys}}
\left(
\tilde{\theta}_i;\mathcal{C}_i
\right),
\label{eq:maml_outer_pinn}
\end{equation}
with $\lambda=1.0$. This formulation allowed the meta-initialization to be shaped by both data fidelity and physical consistency after adaptation, while keeping the inner-loop updates computationally tractable.

The meta-learned initialization was evaluated by fine-tuning on the held-out validation subjects and comparing early adaptation against standard random initialization using the cohort-optimal spectral bias settings.

\section{Results}\label{sec3}

\begin{figure}[!htbp]
\centering
\includegraphics[width=1.0\textwidth]{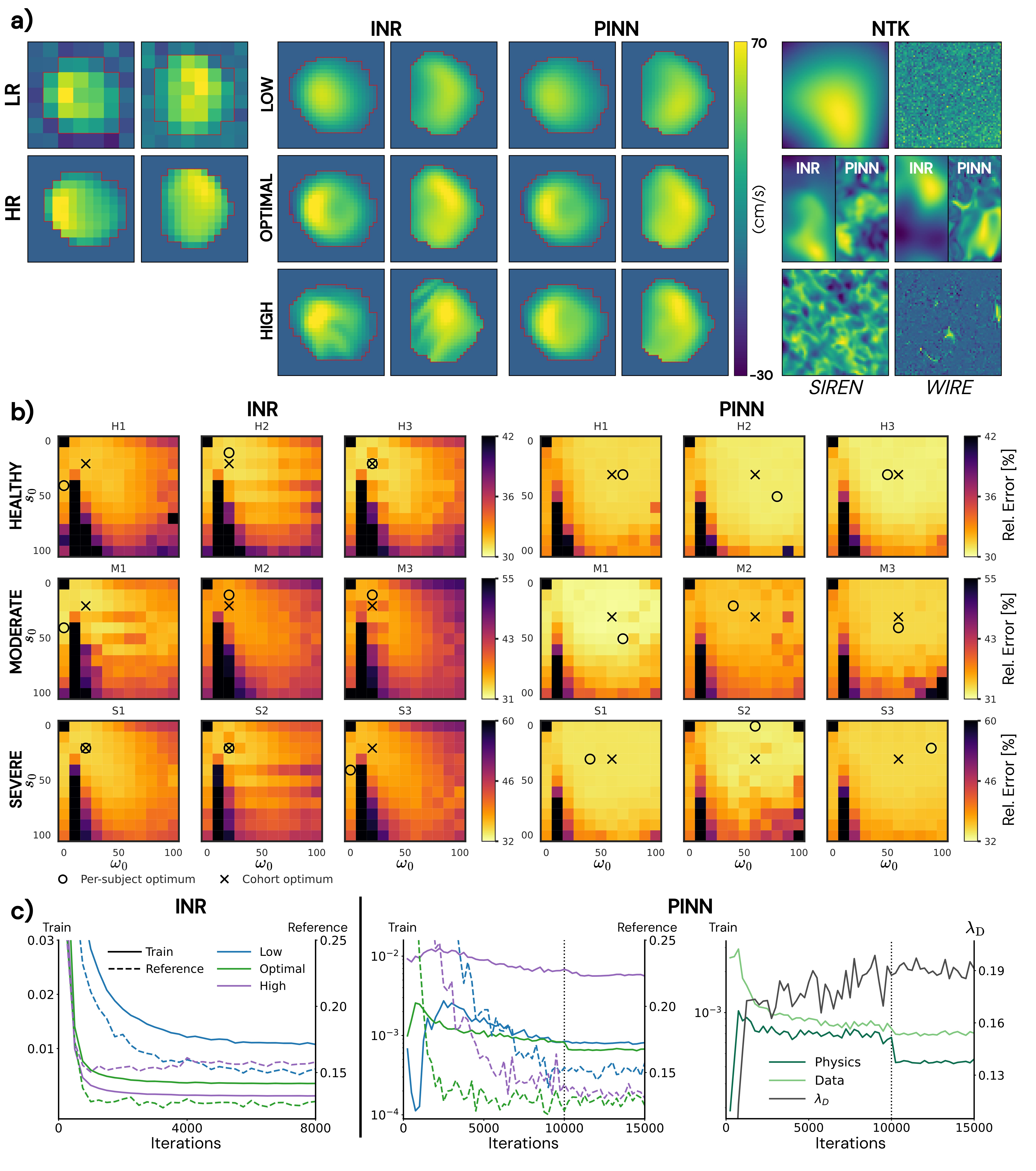}
\caption{\textbf{Spectral bias selection and robustness for data-driven and physics-informed INR.}
\textbf{a)} Representative low-resolution (LR) input and high-resolution (HR) reference slices, reconstructed velocity slices for low ($(\omega_0,s_0)=(5,5)$), optimal ($(20,20)$ for data-driven INR and $(60,30)$ for the physics-informed INR (PINN)), and high ($(90,90)$) spectral bias settings, and neural tangent kernel (NTK) visualizations. The NTK visualizations are generated at initialization for networks using both pure sinusoidal activation functions (SIREN) and Gabor activation functions (WIRE).
\textbf{b)} End-of-training relative error across the $(\omega_0, s_0)$ grid for all subjects, shown separately for the data-driven INR and PINN. Circles indicate subject-specific optima and crosses indicate cohort-optimal settings.
\textbf{c)} Training dynamics for representative low-, optimal-, and high-bias settings. Solid lines denote loss evaluated on the LR training data and dashed lines denote loss against the HR reference. For the physics-informed model, the right panel shows the data loss, physics loss, and the adaptive data-weight $\lambda_D$ for the optimal setting. Vertical dotted lines indicate the transition from Adam to L-BFGS optimization.}
\label{fig3}
\end{figure}

\subsection{Shared spectral bias settings generalize across healthy and diseased cerebrovascular flow}\label{sec31}

We first evaluated how WIRE spectral-bias parameters affected INR reconstruction accuracy and whether optimized settings generalized across subjects. Fig.~\ref{fig3}a shows the influence of WIRE spectral bias across the two implemented implicit models. In a representative case (H1), low spectral bias (Fig.~\ref{fig3}a; top row) produced overly smooth output reconstructions, capturing gross flow profile behavior but failing to recover localized velocity structure visible in the high-resolution (HR) reference. Conversely, high spectral bias (Fig.~\ref{fig3}a; bottom row) increased high-frequency fluctuations, particularly in the data-driven INR. In the middle, and representing the intermediate cohort-optimal settings (Fig.~\ref{fig3}a; middle row; $(\omega_0,s_0) =(20,20)$ for the data-driven INR; $(60,30)$ for the physics-informed INR (the latter using the same coordinate-based architecture but with additional Navier-Stokes regularization, and therefore denoted PINN in the figures)), reconstructions show the ability to preserve HR-like flow structures while still avoiding excessive high-frequency artifacts. To further interpret how different activation settings shape the network's preferred spatial structure before training, we visualized empirical neural tangent kernels (NTKs) at initialization. These visualizations provide a qualitative view of this inductive bias: a pure sinusoidal activation function (SIREN) showed the expected transition from smooth low-frequency modes to finer spatial modes as $\omega_0$ increased, whereas WIRE produced more localized modes reflecting the additional Gaussian envelope controlled by $s_0$.

To assess sensitivity to spectral bias and identify whether settings generalized across anatomies, we systematically varied the WIRE hyperparameters ($(\omega_0,s_0)$, over the range 0-100 with a stride of 10), evaluating end-of-training voxel-wise performance for each configuration. As shown in Fig.~\ref{fig3}b, although subject-specific optima varied across the parameter grid, low-error regions largely overlapped across healthy, moderate, and severe ICAD cases.
This supported the use of shared cohort-optimal settings rather than case-specific tuning, with only a small increase in relative error compared with each subject's own optimum (0.41$\pm$0.60 percentage points for the data-driven INR and 0.26$\pm$0.29 percentage points for the physics-informed INR). Compared with the data-driven INR, the physics-informed INR showed broader low-error regions, suggesting reduced sensitivity to spectral bias choice and a larger set of near-optimal configurations.

The training curves in Fig.~\ref{fig3}c further illustrate why intermediate spectral-bias settings were preferred. For the data-driven INR, the loss for the low bias configuration exhibited a fairly moderate decrease, plateauing at a higher training loss compared to the cohort-optimal setting (final train loss: 0.0108 vs. 0.0036 for the cohort-optimal setting) indicating underfitting of the low-resolution (LR) targets. In contrast, a high bias configuration fitted the underlying data well (final train loss 0.0012), yet its reference loss remained comparatively high (final reference loss: 0.158 vs. 0.128), consistent with limited generalization to the HR reference. The cohort-optimal setting ($(\omega_0, s_0)=(20, 20)$) yielded the lowest reference loss and the smallest train-reference discrepancy, suggesting the most favorable trade-off between data fitting and denoising. For the physics-informed INR, the low bias configuration similarly showed weaker agreement with the HR reference, whereas the difference between the high and optimal bias settings was less pronounced (final reference loss: 0.148 vs. 0.127).

In light of the above, all subsequent experiments used the cohort-optimal settings $(20,20)$ for the data-driven INR and $(60,30)$ for the physics-informed INR.

\subsection{Implicit neural representations recover cerebrovascular flow structure across anatomies and disease severity}\label{sec32}

\begin{figure}[t]
\centering
\includegraphics[width=1.0\textwidth]{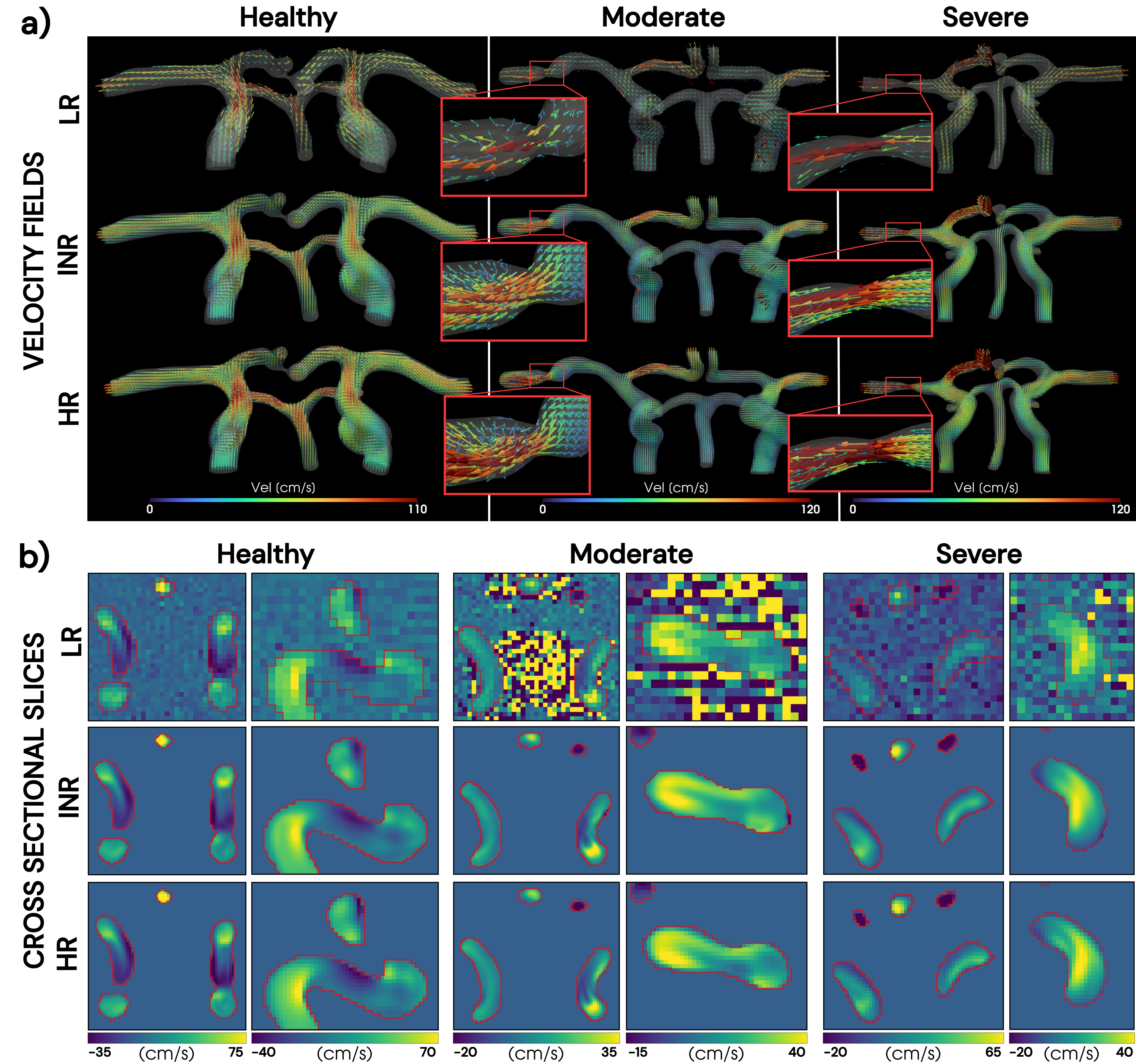}
\caption{\textbf{Representative velocity reconstructions across disease severity.}
\textbf{a)} Three-dimensional velocity magnitude maps at peak systole for representative healthy, moderate ICAD, and severe ICAD cases (H1, M1, and S1), comparing low-resolution (LR) data, data-driven INR reconstruction queried on the HR grid, and high-resolution reference (HR) at peak systole. Insets show magnified regions around the stenosis in the moderate and severe subjects.
\textbf{b)} Representative cross-sectional, uni-directional velocity slices in axial and sagittal orientations for the same cases, with INR reconstructions queried on a grid with two-fold higher spatial sampling than the HR grid.}\label{fig4}
\end{figure}

\begin{table}[t]
\centering
\caption{Performance of data-driven INR and physics-informed INR (PINN) models. Metrics are computed and averaged over the full cardiac cycle, with mean absolute error (MAE), directional error (DE), mean relative error (MRE) and velocity normalized root mean square error (vNRMSE) reported. Bold values indicate the best-performing model for each subject and metric.}
\label{tab:inr_pinn_main}
\setlength{\tabcolsep}{6pt}
\renewcommand{\arraystretch}{1.15}
\begin{tabular}{ll l l S[table-format=1.3] S[table-format=1.3] S[table-format=2.3] S[table-format=1.3]}
\toprule
Severity & Subject & Model & Train time & {MAE [m/s]} & {DE [\%]} & {MRE [\%]} & {vNRMSE} \\
\midrule

\multirow{6}{*}{Healthy}
  & \multirow{2}{*}{H1} & INR  & 1.5 min & \bf{0.111} & 3.5 & 31.52 & 0.084 \\
  &                      & PINN & 126 min & \bf{0.111} & \bf{3.0} & \bf{31.12} & \bf{0.083} \\
  & \multirow{2}{*}{H2} & INR  & 1.7 min & 0.105 & 2.8 & 31.01 & \bf{0.109} \\
  &                      & PINN & 135 min & \bf{0.104} & \bf{2.4} & \bf{30.73} & \bf{0.109} \\
  & \multirow{2}{*}{H3} & INR  &  1.5 min & \bf{0.085} & 2.9 & 30.71 & \bf{0.111} \\
  &                      & PINN & 125 min & \bf{0.085} & \bf{2.6} & \bf{30.66} & \bf{0.111} \\
\midrule

\multirow{6}{*}{Moderate}
  & \multirow{2}{*}{M1} & INR  & 2.0 min & 0.071 & 5.9 & 32.66 & 0.075 \\
  &                      & PINN & 135 min & \bf{0.068} & \bf{4.2} & \bf{31.50} & \bf{0.068} \\
  & \multirow{2}{*}{M2} & INR  & 1.7 min & 0.055 & 3.8 & 35.37 & 0.071 \\
  &                      & PINN & 131 min & \bf{0.053} & \bf{2.9} & \bf{34.28} & \bf{0.069} \\
  & \multirow{2}{*}{M3} & INR  & 1.7 min & 0.141 & 8.6 & 34.65 & 0.062 \\
  &                      & PINN & 137 min & \bf{0.137} & \bf{6.2} & \bf{33.39} & \bf{0.057} \\
\midrule

\multirow{6}{*}{Severe}
  & \multirow{2}{*}{S1} & INR  &  1.6 min   & 0.099 & 7.2 & 34.58 & 0.064 \\
  &                      & PINN & 128 min & \bf{0.098} & \bf{5.2} & \bf{33.88} & \bf{0.062} \\
  & \multirow{2}{*}{S2} & INR  & 2.3 min  & 0.163 & 6.3 & 34.77 & 0.071 \\
  &                      & PINN & 182 min & \bf{0.158} & \bf{4.5} & \bf{33.38} & \bf{0.068} \\
  & \multirow{2}{*}{S3} & INR  & 1.6 min & 0.104 & 6.7 & 36.39 & 0.102 \\
  &                      & PINN & 131 min & \bf{0.100} & \bf{4.1} & \bf{34.78} & \bf{0.092} \\
\bottomrule
\end{tabular}
\end{table}

\begin{table}[t]
\centering
\caption{Performance comparison between data-driven INR and physics-informed INR (PINN) models and CNN-based models (4DFlowNet, GAN-Gen), with tricubic interpolation as a baseline. Metrics are computed voxel-wise over the full cardiac cycle and then averaged across subjects within each severity group, with mean absolute error (MAE), directional error (DE), mean relative error (MRE) and velocity normalized root mean square error (vNRMSE) reported. Bold values indicate the best-performing method within each severity group and metric.}
\label{tab:severity_all_methods_summary}
\setlength{\tabcolsep}{6pt}
\renewcommand{\arraystretch}{1.15}

\begin{tabular}{ll
                l l
                S[table-format=1.3]
                S[table-format=2.1]
                S[table-format=2.3]
                S[table-format=1.3]}
\toprule

Severity & Model & Train time & Inference &
\multicolumn{1}{c}{MAE} [m/s] &
\multicolumn{1}{c}{DE [\%]} &
\multicolumn{1}{c}{MRE [\%]} &
\multicolumn{1}{c}{vNRMSE} \\

\midrule

\multirow{5}{*}{Healthy}
 & Interpolation & -- & -- & 0.117 & 5.1 & 34.64 & 0.128 \\
 & 4DFlowNet           & 30 h & 11.8 s & 0.128 & 6.3 & 36.44 & 0.129 \\
 & GAN-Gen       & 49 h & 12.1 s & 0.112 & 4.0 & 32.71 & 0.127\\
 & INR           & 1.6 min & 1.6 min & \bf{0.103} & 3.2 & 31.19 & 0.097  \\
 & PINN          & 129 min & 129 min & 0.104 & \bf{2.8} & \bf{30.95} & \bf{0.094} \\
\midrule

\multirow{5}{*}{Moderate}
 & Interpolation & - & -     & 0.146 & 10.2 & 41.86 & 0.090  \\
 & 4DFlowNet           & 30 h & 15.0 s & 0.130 & 13.7 & 48.48 & 0.101\\
 & GAN-Gen       & 49 h & 15.0 s & 0.091 & 5.3 & 33.73 & \bf{0.065}  \\
 & INR           & 1.8 min & 1.8 min & 0.082 & 6.0 & 33.60 & 0.071 \\
 & PINN          & 134 min & 134 min  & \bf{0.079} & \bf{4.3} & \bf{32.44} & 0.066 \\
\midrule

\multirow{5}{*}{Severe}
 & Interpolation & - & -     & 0.165 & 8.9 & 41.04 & 0.115\\
 & 4DFlowNet           & 30 h & 14.3 s & 0.176 & 10.9 & 43.70 & 0.107  \\
 & GAN-Gen       & 49 h & 14.2 s & 0.135 & 5.7 & 36.66 & 0.096  \\
 & INR           & 1.8 min & 1.8 min & 0.116 & 6.8 & 35.08 & 0.075  \\
 & PINN          & 147 min & 147 min  & \bf{0.111} & \bf{4.8} & \bf{33.98} & \bf{0.069} \\
\bottomrule
\end{tabular}
\end{table}

Having established shared spectral-bias settings, we next assessed whether the implicit models recovered relevant cerebrovascular flow structure across increasing disease severity. Fig.~\ref{fig4} shows representative reconstructions for one healthy control, one moderate ICAD case, and one severe ICAD case. Qualitatively, the INR reconstructions reproduced the main HR flow patterns, with high and low velocity regions coinciding between the two. For the stenotic cases in particular, magnified views demonstrate improved depiction of localized high-velocity flow near the narrowed vessel segments, a region in which low-resolution imaging failed to provide any comprehensive visual overview of the transstenotic flows. Cross-sectional slices (Fig.~\ref{fig4}b) also showed reduced voxel-level noise and close agreement with the HR reference (additional full-field visualizations for the remaining subjects are provided in Supplementary Fig. S1).

Complementing the qualitative findings, quantitative subject-level performance for the data-driven INR and physics-informed INR is summarized in Table~\ref{tab:inr_pinn_main}. As shown, comparably low errors are observed across all models and network configurations (cohort-averaged vNRMSE: 0.083 for the data-driven INR and 0.080 for the physics-informed INR),
indicating robust performance using the pre-selected cohort-optimal spectral-bias settings. Further, across all subjects, the physics-informed INR achieved the strongest quantitative performance, with average DE decreasing from 5.3\% to 3.9\% and MRE decreasing from 33.5\% to 32.6\% compared with the data-driven INR. However, these gains were modest in absolute magnitude, with cohort-averaged MAE differing by 0.2 cm/s (10.4 vs. 10.2 cm/s). The training time differed substantially between approaches, with the data-driven INR converging within 1.5-2.3 minutes per subject, whereas the physics-informed INR required 125-182 minutes.

Table~\ref{tab:severity_all_methods_summary} compares the implicit models to tricubic interpolation and reference supervised CNN-based baselines, averaged within each severity group. The INR setups generally outperform alternative approaches, with data-driven INR error metrics improving as compared to tricubic interpolation (cohort-averaged MAE: 0.100 vs 0.143, vNRMSE: 0.081 vs 0.111), 4DFlowNet (MAE: 0.145, vNRMSE: 0.112) and GAN-Gen (MAE: 0.113, vNRMSE: 0.096), respectively. Notably, differences between the data-driven INR and the alternative approaches were particularly pronounced in the stenotic groups (MAE improvement in healthy, moderate, and severe cases: 2.5/4.8/6.0 cm/s vs. 4DFlowNet), highlighting the potential of the coordinate-based networks in pathophysiological flow settings. Lastly, Table~\ref{tab:severity_all_methods_summary} also highlights fundamental computational trade-offs between approaches: the CNN setups required several hours of training, but once trained, inference required only 12-15 s. The implicit models on the other hand required re-training for each subject, with average training times of 1.6-1.8 min for the data-driven, and 129-147 min for the physics-informed INR, respectively.

\subsection{Implicit neural representations improve robustness to noise and VENC-dependent aliasing}\label{sec33}

\begin{figure}[t!]
\centering
\includegraphics[width=1.0\textwidth]{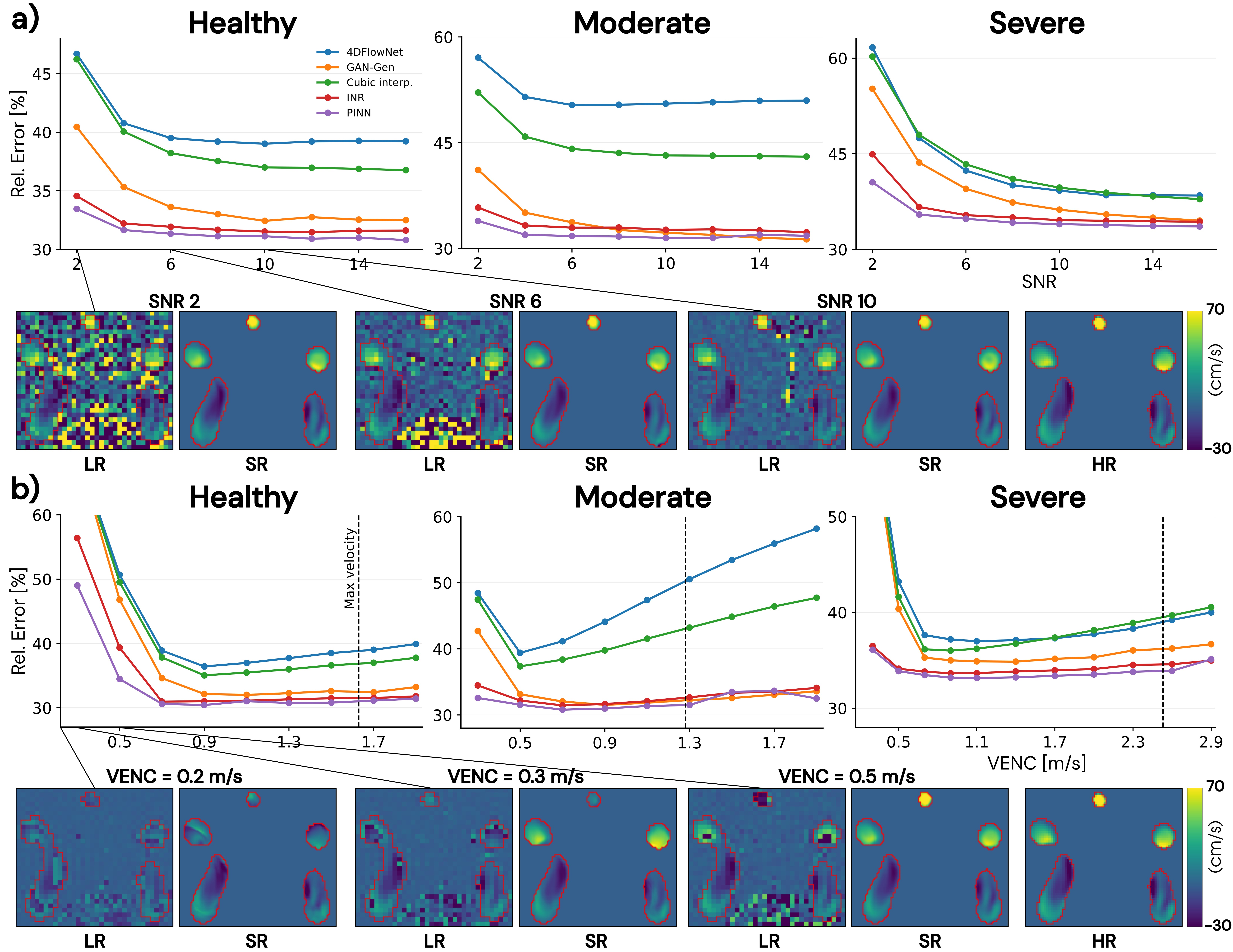}
\caption{\textbf{Robustness to noise and VENC-dependent aliasing.}
\textbf{a)} Relative error as a function of added measurement noise, expressed by signal-to-noise ratio (SNR), for representative healthy, moderate ICAD, and severe ICAD cases (H1, M1, and S1). Curves compare 4DFlowNet, GAN-Gen, tricubic interpolation, data-driven INR, and physics-informed INR (PINN). Representative cross-sectional velocity slices for H1 show LR and data-driven INR reconstructions (SR) at SNR = 2, 6, and 10, with the HR reference shown for comparison.
\textbf{b)} Relative error as a function of velocity encoding (VENC) for the same representative cases. Dashed vertical lines indicate the maximum velocity in each case. Representative cross-sectional velocity slices for H1 show LR and data-driven INR reconstructions (SR) at VENC = 0.2, 0.3, and 0.5 m/s.
}\label{fig5}
\end{figure}

We next evaluated reconstruction robustness under acquisition degradations introduced during synthetic data generation to mimic realistic image quality. As shown in Fig.~\ref{fig5}a (top row), increasing SNR reduced relative error for all methods, reflecting the expected improvement with higher input data quality. At an SNR representative of a good-quality clinical acquisition (SNR=10), the physics-informed INR achieved the lowest average relative error (32.2\%), closely followed by the data-driven INR (32.9\%) and GAN-Gen (33.6\%), while interpolation and 4DFlowNet showed higher errors (40.0\% and 42.9\%, respectively). Notably, the advantage of the implicit models became more pronounced under high noise configurations: at SNR=2, the physics-informed INR and data-driven INR achieved average errors of 36.0\% and 38.4\%, compared with 45.6\% for GAN-Gen, 52.9\% for interpolation, and 55.1\% for 4DFlowNet. This behavior is qualitatively illustrated by the representative axial slices in Fig.~\ref{fig5}a (bottom row), shown at the level of the internal carotid arteries (ICAs) and basilar artery (BA) during peak systole. Across representative and increasingly degraded noise conditions (SNR = 10, 6, and 2), the data-driven INR reconstruction suppressed voxel-level noise and recovered velocity patterns that closely resembled the HR reference.

The acquisition-specific VENC setting also directly affects image quality. If VENC is set too low, velocities above the VENC appear as wrapped artificial velocity signals; if VENC is set too high, velocity-to-noise ratio decreases because the same phase noise corresponds to larger velocity uncertainty. The VENC curves in Fig.~\ref{fig5}b (top row) reflected this trade-off, showing a U-shaped behavior: errors increased at low VENC due to phase wrapping, and again at high VENC due to reduced phase sensitivity. Here, the implicit models showed lower and flatter error curves than interpolation and the CNN-based baselines, indicating greater tolerance to both aliasing and high-VENC noise amplification.
This was particularly evident at low VENC settings: at VENC = 0.3 m/s, the data-driven and physics-informed INR achieved errors of 34.5\% and 32.6\% in the moderate ICAD case, and 36.5\% and 36.1\% in the severe ICAD case, respectively, remaining substantially below the corresponding errors of interpolation, 4DFlowNet, and GAN-Gen. The representative cross-sectional slices in Fig.~\ref{fig5}b (bottom row) illustrate this effect qualitatively. At VENC = 0.5 m/s, the data-driven INR corrected phase-wrapping artifacts present in the LR input. At VENC = 0.3 and 0.2 m/s, aliasing became more severe, and aliasing-related errors remained in the reconstructed slices.

\subsection{Meta-learned initialization accelerates patient-specific adaptation}\label{sec34}

\begin{figure}[t!]
\centering
\includegraphics[width=1.0\textwidth]{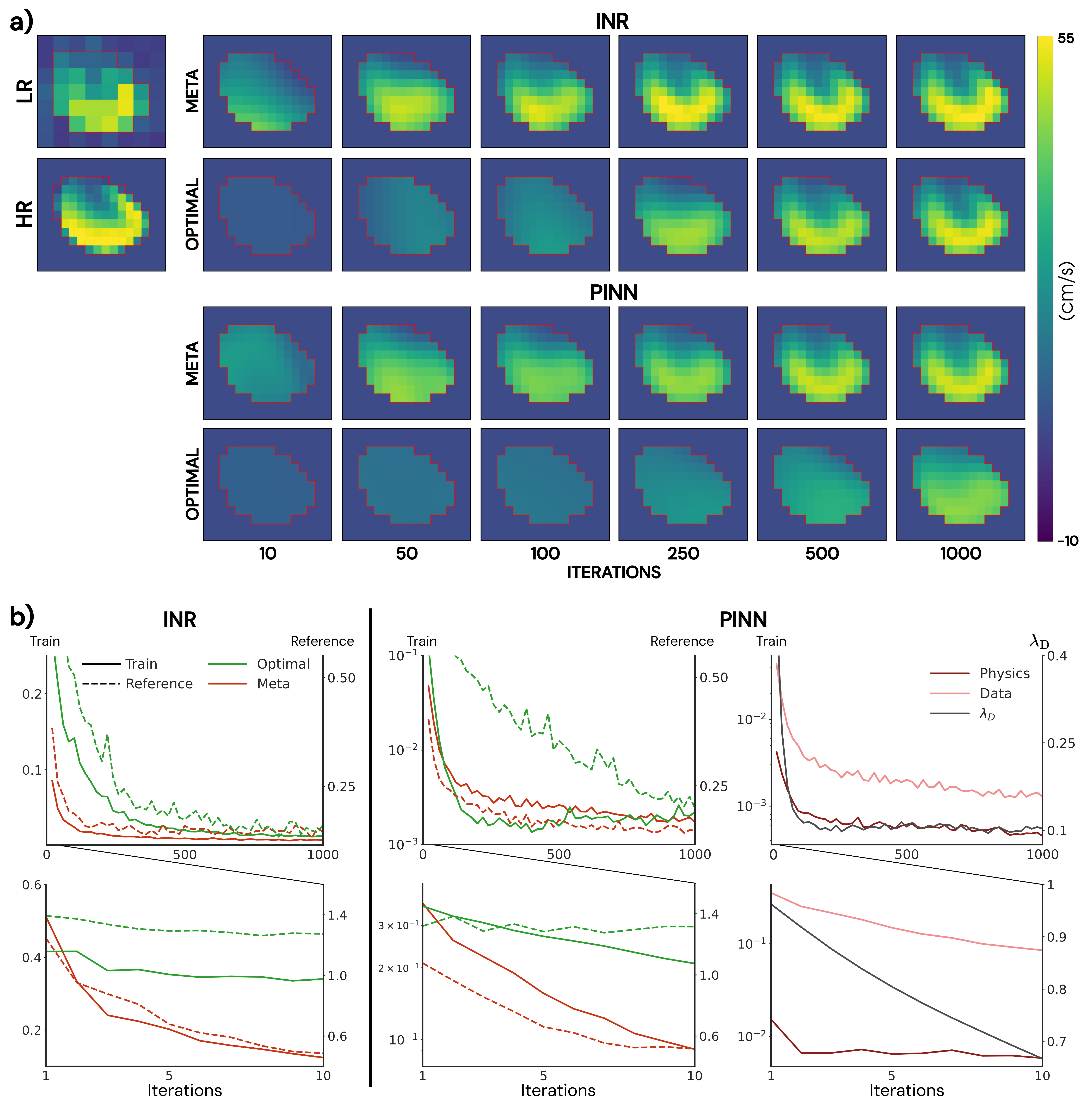}
\caption{
\textbf{Meta-learned initialization accelerates early adaptation.}
\textbf{a)} Cross-sectional velocity reconstructions for a held-out validation (H3) subject after 10, 50, 100, 250, 500, and 1000 fine-tuning iterations. Results are shown for the data-driven INR and physics-informed INR (PINN), comparing meta-learned initialization (META) with standard initialization using the cohort-optimal spectral bias setting (OPTIMAL). Low-resolution input (LR) and high-resolution reference (HR) are shown for comparison.
\textbf{b)} Training dynamics for the same validation subject over the first 1000 iterations, with zoomed views over the first 10 iterations. Solid lines denote loss evaluated on the LR training data and dashed lines denote loss against the HR reference. For the PINN, data loss, physics loss, and adaptive data-weight $\lambda_D$ are also shown for the corresponding iterations.}
\label{fig6}
\end{figure}

\begin{table}[t]
\centering
\caption{Performance comparison between meta-learned and standard initialization at early training.
Metrics are reported for each held-out validation subject (H3, M3, S3) after 500 iterations for the data-driven INR and the physics-informed INR (PINN). Metrics are computed voxel-wise over the full cardiac cycle, with mean absolute error (MAE), directional error (DE), mean relative error (MRE) and velocity normalized root mean square error (vNRMSE) reported. META denotes fine-tuning from the meta-learned initialization, and OPTIMAL denotes standard initialization using the cohort-optimal spectral-bias setting. Bold values indicate the best-performing method within each severity group, INR model and error metric.}
\label{tab:meta_learning_500}
\setlength{\tabcolsep}{5pt}
\renewcommand{\arraystretch}{1.15}
\begin{tabular}{ll l c c c c c}
\toprule
Subject & Model & Initialization & Train time & MAE [m/s] & DE [\%] & MRE [\%] & vNRMSE \\
\midrule
\multirow{4}{*}{H3}
& \multirow{2}{*}{INR}  & OPTIMAL  & 0.5 min  & 0.093 & \bf{2.6} & 32.39 & 0.122 \\
&                       & META     & 0.5 min & \bf{0.089} & 3.0 & \bf{31.70} & \bf{0.116} \\
& \multirow{2}{*}{PINN} & OPTIMAL  & 4.2 min & 0.146 & 5.4 & 43.49 & 0.183 \\
&                       & META     & 4.2 min & \bf{0.095} & \bf{3.1} & \bf{32.69} & \bf{0.128} \\
\midrule
\multirow{4}{*}{M3}
& \multirow{2}{*}{INR}  & OPTIMAL  & 0.5 min & 0.162 & \bf{8.0} & 38.06 & 0.070 \\
&                       & META     & 0.5 min & \bf{0.144} & 9.1 & \bf{35.33} & \bf{0.066} \\
& \multirow{2}{*}{PINN} & OPTIMAL  & 4.2 min & 0.201 & 7.5 & 42.64 & 0.082 \\
&                       & META     & 4.2 min & \bf{0.153} & \bf{6.8} & \bf{36.05} & \bf{0.068} \\
\midrule
\multirow{4}{*}{S3}
& \multirow{2}{*}{INR}  & OPTIMAL  & 0.5 min & 0.115 & 8.3 & 38.33 & 0.114 \\
&                       & META     & 0.5 min & \bf{0.101} & \bf{6.0} & \bf{35.04} & \bf{0.103} \\
& \multirow{2}{*}{PINN} & OPTIMAL  & 4.2 min & 0.155 & 5.5 & 45.18 & 0.136 \\
&                       & META     & 4.2 min & \bf{0.111} & \bf{4.3} & \bf{37.36} & \bf{0.106} \\
\bottomrule
\end{tabular}
\end{table}

Since INRs are trained separately for each subject, reconstruction requires patient-specific training at inference time. While this provides flexibility and avoids the need for paired supervised training data, it may limit practical use if optimization is too slow. For this reason, we next evaluated whether so-called \emph{meta-learning} could accelerate patient-specific adaptation. Utilizing a model-agnostic meta-learning approach as introduced by Finn et al.~\cite{finn2017model}, the setup seeks to learn a shared initialization applicable \emph{across} training subjects with continuous fine-tuning performed from this meta-learned initialization stage, as compared to a standardized re-training approach. Evaluation was performed on held-out validation subjects not used during meta-training, with implementation details provided in Section~\ref{sec213}.

Fig.~\ref{fig6}a shows representative reconstructions during early fine-tuning. For the data-driven INR, the meta-learned initialization produced anatomically plausible velocity structure within the first tens of iterations and closely approached the HR reference by approximately 100 iterations. In contrast, the standard initialization evolved more gradually and required substantially more iterations to recover comparable localized flow structure. A similar trend was observed for the physics-informed INR. Meta-learned initialization produced more coherent reconstructions throughout early adaptation, whereas standard initialization remained visually underdeveloped over the first 1000 iterations. However, adaptation of the physics-informed model remained slower than for the data-driven INR, consistent with the added complexity of jointly optimizing data fidelity and physics-based regularization.

The corresponding loss curves further confirmed accelerated adaptation (Fig.~\ref{fig6}b). For both the data-driven and physics-informed INR, meta-learned initialization reduced early training and reference losses compared with standard initialization. The advantage was most pronounced during the first few iterations, where the meta-initialized models started closer to a subject-specific solution and rapidly recovered the dominant velocity structure. Quantitative evaluation after 500 iterations supported these findings (Table~\ref{tab:meta_learning_500}). Averaged over the three held-out validation subjects, meta-learned initialization reduced MAE from 0.123 to 0.111 for the data-driven INR and from 0.167 to 0.120 for the physics-informed INR, with corresponding reductions in vNRMSE from 0.102 to 0.095 and from 0.134 to 0.101, respectively. Together, these results indicate that meta-learned initialization improves early patient-specific adaptation, with particularly strong potential benefit for the computationally heavier physics-informed model.

\subsection{Implicit neural representations enable direct in-vivo cerebrovascular reconstruction}\label{sec35}

\begin{figure}[t!]
\centering
\includegraphics[width=1.0\textwidth]{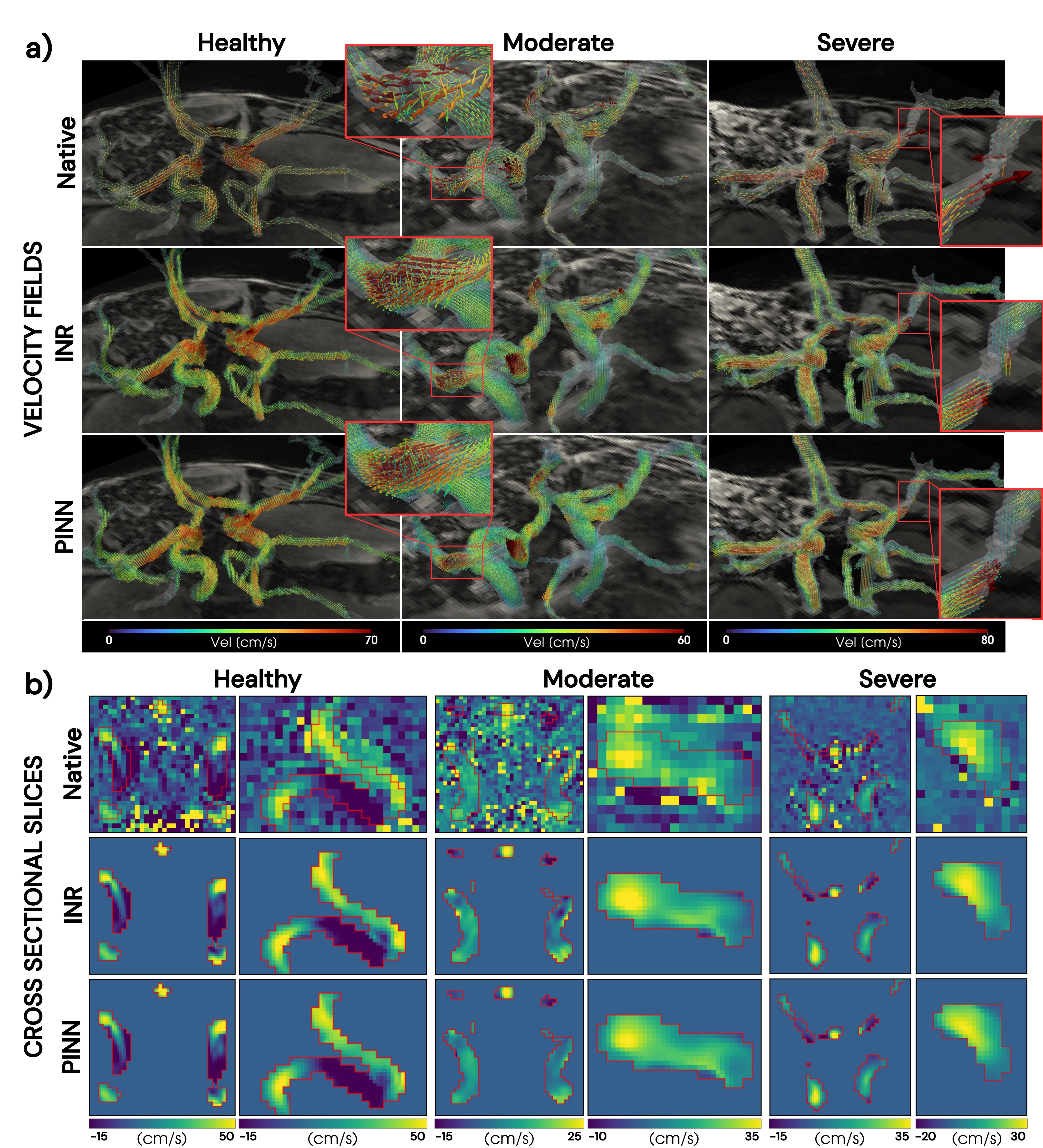}
\caption{\textbf{Qualitative in-vivo reconstructions across disease severity.}
\textbf{a)} Three-dimensional velocity magnitude maps at peak systole for representative healthy, moderate ICAD, and severe ICAD cases (H1, M1, and S1), comparing acquired native 4D Flow MRI data with direct \textit{in-vivo} reconstructions from the data-driven INR and physics-informed INR (PINN) at peak systole. Insets show magnified regions around the stenosis
in the moderate and severe subjects.
\textbf{b)} Representative cross-sectional, uni-directional velocity slices in axial and sagittal orientations for the same cases, with INR reconstructions queried on a grid with two-fold higher spatial sampling than the native grid.}
\label{fig7}
\end{figure}

Fig.~\ref{fig7} presents qualitative \textit{in-vivo} reconstructions obtained by fitting the implicit models directly to acquired cerebrovascular 4D Flow MRI data in representative healthy, moderate ICAD, and severe ICAD cases. In all three cases, both the data-driven and physics-informed INR produced velocity fields that were visually smoother and more spatially coherent than the native input, consistent with reduced voxel-level noise and preservation of the dominant vascular flow patterns. This behavior is evident in both the three-dimensional velocity maps and the representative cross-sectional slices.

In the moderate ICAD case, the reconstructions showed coherent post-stenotic flow characteristics, with less fragmented vector patterns and clearer localized high-velocity structure downstream of the narrowing. In the severe ICAD case, the implicit models similarly improved flow coherence in the larger surrounding vessels. However, velocity recovery within the narrowed, stenotic segment remained limited, as a result of the poor coverage of the native low-resolution data in this region. Thus, while the implicit models reduced noise and improved spatial coherence, they did not infer reliable flow structure where the acquired data provided insufficient information content.

Additional full-field visualizations for the remaining subjects are provided in Supplementary Fig. S2. Because no noise-free, high-resolution \textit{in-vivo} references are available, this analysis remained qualitative.

\section{Discussion}\label{sec4}

In this study, we introduced an INR framework for unsupervised denoising and super-resolution of intracranial 4D Flow MRI. Using subject-specific synthetic datasets across controlled low- and high-resolution settings, we showed that both data-driven and physics-informed INRs improved velocity reconstruction across healthy and stenotic vessel configurations. The physics-informed INR achieved the strongest quantitative performance, with modest but consistent gains over the data-driven equivalent. However, this improvement came at a substantial computational cost, where the data-driven INR achieved near-comparable accuracy at a fraction of the required training time. This extends prior work by evaluating continuous, patient-specific reconstruction in complex cerebrovascular anatomy, including systematic spectral-bias analysis, meta-learning attempts, and direct fitting to acquired in-vivo 4D Flow MRI data. Together, these findings support INRs as practical and flexible models for improving quantitative cerebrovascular 4D Flow MRI under constrained acquisition conditions.

A central finding was that spectral bias strongly influenced INR reconstruction quality (Section~\ref{sec31}). Low spectral bias settings produced overly smooth velocity fields, capturing gross vascular structure but failing to recover localized flow features. Conversely, high spectral bias increased sensitivity to high-frequency fluctuations and noise-like patterns. Intermediate, optimal settings provided a sought-after balance between detail recovery and denoising. Herein, optimal spectral bias settings and associated regions of optimal performance overlapped substantially across subjects and disease severity groups, indicating that a shared cohort-level setting can be used without per-subject hyperparameter tuning. This is an important practical result, since extensive case-specific tuning limits clinical applicability and hinders generalized network usage. The physics-informed INR showed even broader low-error regions, suggesting that an included physics-based regularization conformed output to realistic flow phenomena and further reduced sensitivity to individual spectral bias choices.
The physics-informed INR also achieved the strongest quantitative performance; however, these gains were modest in absolute magnitude and came at substantially higher computational cost. This suggests that much of the reconstruction improvement is attributable to the coordinate-based INR and its optimized spectral-bias configuration, while added physics-informed regularization provides incremental accuracy gains at the expense of longer optimization.

The robustness experiments (Section~\ref{sec33}) further demonstrated the value of implicit reconstruction under degraded, but realistic, acquisition conditions. Increasing SNR improved performance for all methods, but the implicit models maintained lower relative error than alternative reference approaches, particularly under high-noise conditions. VENC-specific experiments showed the expected trade-off between phase wrapping at low VENC and reduced phase sensitivity at high VENC, where the periodic anti-aliasing loss improved tolerance to moderate phase wrapping as compared to non-optimized interpolation and CNN-based methods. However, this tolerance was not unlimited:
aliasing-related errors remained when the VENC was too low (Fig.~\ref{fig5}),
highlighting that the model cannot fully recover reliable velocity information when the acquired phase data are too strongly corrupted.

The clinical relevance of this work lies in addressing a key barrier for cerebrovascular 4D Flow MRI: the inherent trade-off between spatiotemporal resolution, acquisition time, and image noise. In the intracranial circulation, small vessels, complex anatomies, and possible stenotic flow accelerations exacerbate these trade-offs; a matter of particular concern where hemodynamic data quantification has shown direct dependence on image quality and effective resolution~\cite{marlevi2021noninvasive, aristova2022accelerated}. Improved velocity reconstruction may increase confidence in downstream hemodynamic analyses, including flow distribution, collateral flow assessment, stenosis characterization, and pressure-related or wall-based biomarkers. With this noted, the present results should be interpreted as a step \emph{toward} improved quantitative imaging rather than evidence of clinical utility. For the latter, future work should evaluate the clinical added value of improved voxel-wise velocity fidelity as introduced by an implicit network reconstruction, continuing on the systematic quantification presented in a controlled environment in our current study.

Our work introduces several specific novelties compared with existing state-of-the-art approaches for 4D Flow MRI super-resolution. CNN-based methods have been shown to improve spatial resolution, reduce noise, or recover flow features in standard clinical data as exemplified by Rutkowski et al.~\cite{rutkowski2021enhancement}, 4DFlowNet~\cite{ferdian20204dflownet}, or SRFlow~\cite{shit2022srflow}. Being supervised setups, however, these approaches require comparably large and diverse datasets to generalize across wider anatomical setups. Furthermore, conventional CNNs operate on fixed, voxelized grids, limiting their ability to represent continuous spatial fields or reconstruct data at arbitrary spatial scales. In contrast, the present framework is optimized directly on the measured velocity field of each subject, requiring no supervised paired training data, and representing the reconstruction as a continuous coordinate-based function.

This work also differs from prior PINN applications in hemodynamic imaging. Previous studies have demonstrated the value of physics-informed learning for vascular flow reconstruction and pressure inference, including applications to aortic pressure estimation, aneurysmal flow, ventricular flow, and 4D Flow MRI reconstruction~\cite{kissas2020machine, fathi2020super, arzani2021uncovering, zhang2023physics, shone2023deep, kang2025flow, bisbal2026unsupervised}, however, extensions to clinically acquired intracranial data remain scarce. Herein, these restrictions arise because many implementations either depend directly on underlying CFD simulations and their associated meshes, limiting application to voxelized image data~\cite{arzani2021uncovering, zhang2023physics, shone2023deep}, or use simplified formulations of the Navier-Stokes equations for computational efficiency (one-dimensionality, axisymmetric assumptions, etc.~\cite{kissas2020machine, sarabian2022physics}). Among the studies that employ a full three-dimensional Navier-Stokes description, efforts remain limited to idealized \textit{in-silico} configurations~\cite{fathi2020super, shone2023deep}, leaving clinical utility unknown. As such, our work provides novel insights and utility for INR adoption, reconstructing 3D velocity fields voxel-by-voxel in tortuous, multi-bifurcated vascular compartments.

The closest prior INR study in 4D Flow MRI is the work by Saitta et al.~\cite{saitta2024implicit}, who demonstrated rapid unsupervised super-resolution and denoising in a localized aortic setting using a SIREN-based representation. Our study extends this line of work in several important ways. First, we evaluate INR reconstruction in a more complex vascular territory, with the neurovasculature characterized by small, tortuous, and bifurcating vessels. Second, we systematically characterize the dependence of reconstruction accuracy on spectral bias, showing how the frequency response of the representation governs the trade-off between fine-scale recovery and noise suppression. Third, we compare a fast data-driven formulation with a physics-informed extension. Together, these elements move INR-based 4D Flow MRI reconstruction beyond proof-of-concept super-resolution and toward practical questions in cerebrovascular imaging: how the representation should be tuned, to what extent physics-informed regularization is beneficial, and how robust the method is under realistic acquisition degradations.

A few limitations should be acknowledged. First, the cohort included nine subjects, limiting the ability to draw population-level conclusions about performance across the full range of cerebrovascular anatomy and pathology. However, the cases were deliberately selected to span healthy to severe ICAD anatomies, thereby covering a clinically relevant range of intracranial flows. Moreover, the proposed framework is optimized directly on each subject rather than trained as a fully supervised population model, reducing the dependence on cohort size for the main experiments. Second, the relevance of the quantitative evaluation depends on the realism of the synthetic 4D Flow MRI generation. Although the downsampling pipeline was designed to mimic key acquisition effects, including magnitude-informed complex image generation, k-space cropping, VENC-dependent phase encoding, and realistic noise injection~\cite{odeback2026potential}, it cannot fully capture all sources of \textit{in-vivo} variability, such as coil-sensitivity effects, patient motion, reconstruction artifacts, or pathological changes in magnitude signal. The purpose of the synthetic benchmark was therefore not to perfectly reproduce clinical acquisition, but to provide a controlled and clinically plausible setting for testing whether the proposed reconstructions improve velocity fields sufficiently to support downstream hemodynamic analysis. Third, the physics-informed INR remains computationally expensive, limiting its immediate practicality for time-sensitive workflows. Herein, both INR variants may benefit from further systematic optimization of training choices, including loss weighting, collocation sampling, spectral bias settings, learning-rate schedules, and optimizer configuration. Such optimization could further incrementally improve convergence speed, stability, and reconstruction accuracy.
Fourth, while quantitative metrics were explored, the effects of resolution enhancement and concurrent denoising remain intertwined. Our results should therefore be viewed as an assessment of the \emph{combined} effect of enhancing image quality in terms of both signal quality and recovered resolution, with alternative metrics assessing noise and resolution effects separately \cite{jacobs2026fourier} to be explored in future, separate work. Finally, comparison with supervised CNN baselines should be interpreted in the context of training-domain differences and implementation constraints; nevertheless, the results highlight the domain-specificity of supervised approaches and the potential flexibility of patient-specific implicit reconstruction.

In summary, INRs provide a promising foundation for robust cerebrovascular 4D Flow MRI enhancement. The data-driven INR offers fast, unsupervised patient-specific reconstruction, while the physics-informed variant provides modest but consistent accuracy gains and improved physical regularization at higher computational cost. By combining continuous neural field representations, spectral bias control, anti-aliasing losses, and physics-informed constraints, this framework provides a flexible route toward more reliable quantitative cerebrovascular flow imaging under realistic and diverse acquisition constraints.

\section{Data availability}\label{sec5}

The datasets generated and analyzed during the current study are not publicly available due to institutional and ethical restrictions on sharing imaging data derived from human participants. Processed and synthetic derivatives supporting the findings of this study may be made available from the corresponding author upon reasonable request, subject to institutional approvals and data-sharing agreements.

\section{Code availability}

The code used for data-driven and physics-informed INR training, synthetic 4D Flow MRI data generation, evaluation, and visualization is publicly available at \url{https://github.com/oliverwelinodeback/4DFlowINR}.

\section{Author contributions}

O.W.O.: conceptualization, implementation, methodology, analysis, evaluation, writing - original draft. N.G.A.: methodology, computational modelling, writing - review \& editing. S.J.: implementation, methodology, analysis, writing - review \& editing. J.B.: implementation, methodology, writing - review \& editing. E.F.: early idea discussion and conceptualization, methodology, writing - review \& editing. A.A.Y.: early idea discussion and conceptualization, writing - review \& editing. P.C.: methodology, writing - review \& editing. S.A.A.: data acquisition, writing - review \& editing. M.B.: methodology, computational modelling, writing - review \& editing. D.N.: methodology, supervision, writing - review \& editing. P.W.: data acquisition, writing - review \& editing. S.S.: data acquisition, supervision, writing - review \& editing. T.G.: supervision, analysis, writing - review \& editing. A.F.: methodology, analysis, supervision, writing - review \& editing. D.M.: conceptualization, methodology, analysis, evaluation, writing - original, review \& editing. All authors have read and approved the manuscript.

\section{Acknowledgements}

Computations were performed on resources provided by the National Academic Infrastructure for Supercomputing in Sweden at the National Supercomputer Centre at Linköping University (Berzelius).

This work was funded in part by the European Union (ERC, MultiPRESS, 101075494) and by the National Heart, Lung, and Blood Institute of the National Institutes of Health (research grant: 1R01HL149787). Views and opinions expressed are those of the authors and do not reflect those of the European Union, the European Research Council Executive Agency, the National Institutes of Health, or the National Heart, Lung, and Blood Institute.

\section{Competing interests}

All authors declare no financial or non-financial competing interests.

\section{Declaration of generative AI and AI-assisted technologies in the writing process}

During the preparation of this work, the authors used ChatGPT (OpenAI) to assist in improving the readability and language of the manuscript. After using this tool, the authors reviewed and edited the content as needed and take full responsibility for the content of the published article.

\bibliography{sn-bibliography}

\clearpage
\section*{Supplementary Material}
\setcounter{section}{0}
\setcounter{figure}{0}
\setcounter{table}{0}
\setcounter{equation}{0}
\renewcommand{\thesection}{S\arabic{section}}
\renewcommand{\thefigure}{S\arabic{figure}}
\renewcommand{\thetable}{S\arabic{table}}
\renewcommand{\theequation}{S\arabic{equation}}

\section{Additional reconstruction examples}

\begin{figure}[t]
  \centering
  \includegraphics[width=\linewidth]{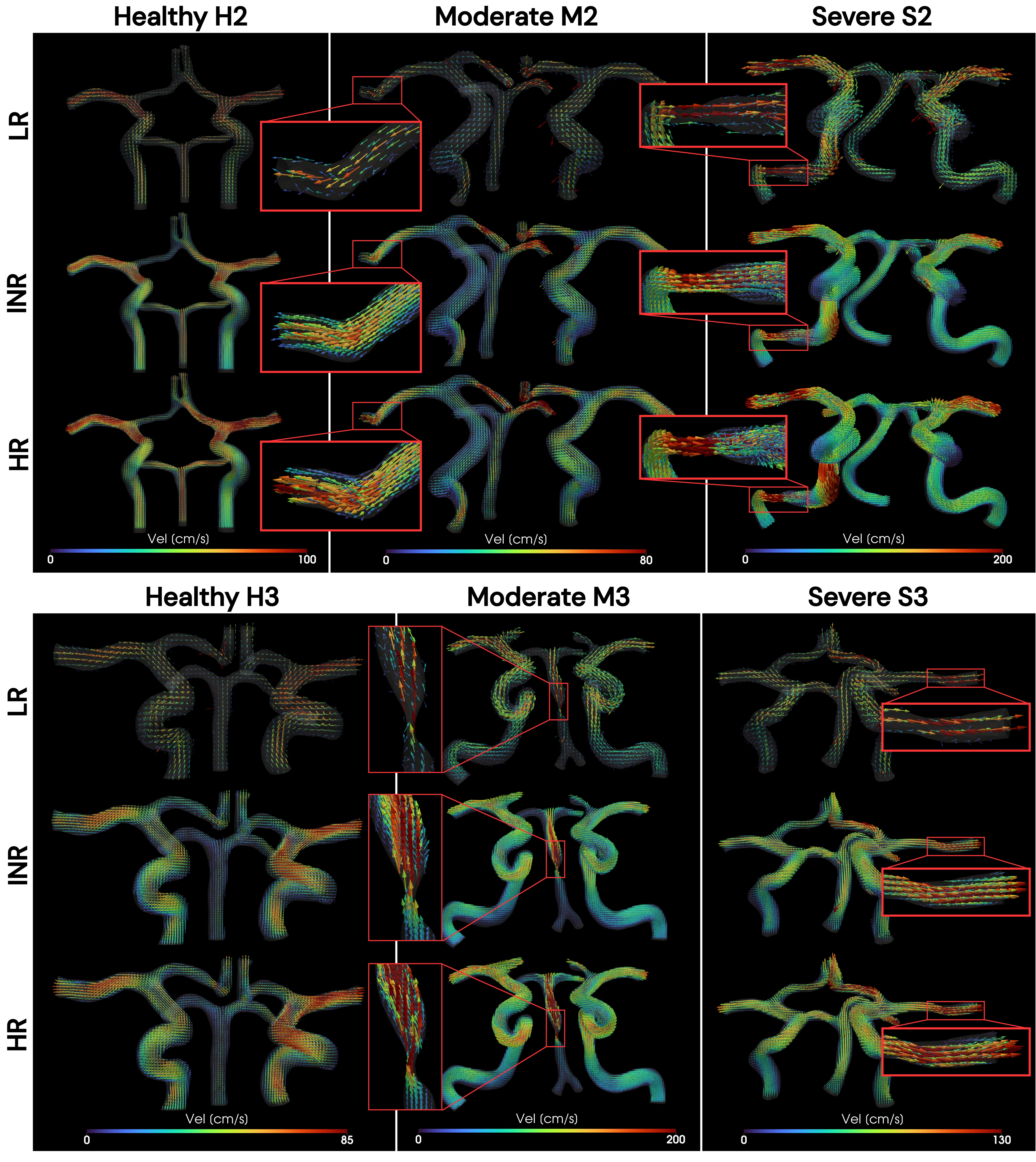}
  \caption{\textbf{Additional in-silico velocity reconstructions across disease severity.}
  Three-dimensional velocity magnitude maps at peak systole for the remaining healthy, moderate ICAD, and severe ICAD cases, comparing low-resolution data, data-driven INR reconstruction queried on the HR grid, and high-resolution reference (HR) at peak systole. Insets show magnified regions around the stenosis in the moderate and severe subjects.}
  \label{SFig1}
\end{figure}

\begin{figure}[t]
  \centering
  \includegraphics[width=\linewidth]{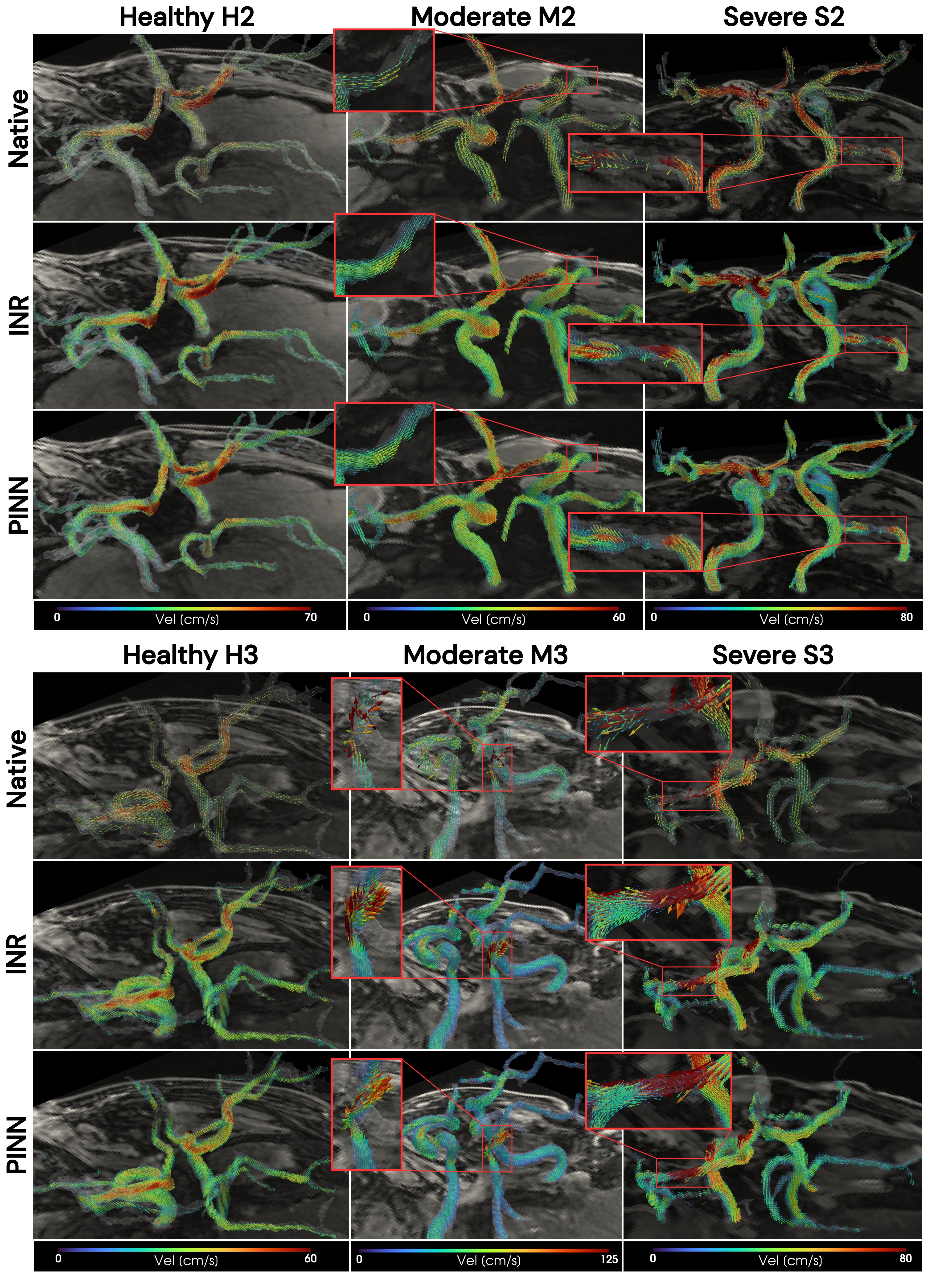}
  \caption{\textbf{Additional in-vivo reconstructions across disease severity.}
  Three-dimensional velocity magnitude maps at peak systole for the remaining healthy, moderate ICAD, and severe ICAD cases, comparing acquired native 4D Flow MRI data with direct in-vivo reconstructions from the data-driven INR and physics-informed INR (PINN) at peak systole. Insets show magnified regions around the stenosis in the moderate and severe subjects.}
  \label{SFig2}
\end{figure}

Supplementary Figs.~\ref{SFig1} and~\ref{SFig2} provide additional full-field visualizations for subjects not shown in the main text. These examples complement the representative in-silico and in-vivo reconstructions shown in Figs.~4 and~7 of the main manuscript.

\end{document}